\documentclass[%
preprint, tightenlines, 
preprintnumbers,
nofootinbib,
 amsmath,amssymb,
 aps,
 showkeys,
 showpacs,
]{revtex4-2}

\usepackage{graphicx}
\usepackage{dcolumn}
\usepackage{bm}
\usepackage{slashed} 

\begin{document}
\renewcommand{\arraystretch}{1.2}

\title{Chiral Quark Soliton Model and Nucleon Parton Distribution~Functions}


\author{Masashi Wakamatsu}
\affiliation{%
KEK Theory Center, Institute of Particle and Nuclear Studies,
High Energy Accelerator Research Organization (KEK),
Oho 1-1, Tsukuba, 305-0801, Ibaraki, Japan
}%




\date{\today}

\begin{abstract}
The chiral quark soliton model (CQSM) is an effective quark model of baryons
maximally taking account of the most important feature of low-energy QCD, i.e.,
the spontaneous chiral symmetry breaking of the QCD vacuum and the associated 
appearance of Nambu--Goldstone pions. 
It shares many common features with the famous Skyrme model in that
the baryons are viewed as rotating hedgehog objects in both models. 
Despite many similarities, it turned out that the CQSM can give more 
realistic predictions on most baryon observables. 
Above all, a decisive advantage of the CQSM over the Skyrme-like models is that 
it can handle non-local quark--quark correlations in baryons, which is absolutely 
impossible within the framework of effective meson theories.
This feature is decisively important for making theoretical predictions on the quark 
distribution functions inside the nucleon, which are defined as nucleon matrix elements 
of bilinear quark operators with light-cone separation.
In the present paper, we try to elucidate why and how the CQSM can give successful 
predictions for a variety of types of nucleon quark distribution functions, especially for 
the flavor asymmetry of the unpolarized and longitudinally polarized 
sea-quark (anti-quark) distribution functions in the nucleon.
\end{abstract}


\vspace{3mm}

\vspace{3mm}




\maketitle
\newpage


\section{Introduction}
\label{sec:introduction}

It is widely believed that the most important properties of the quantum 
chromodynamics (QCD) are the color confinement and the asymptotic freedom.
However, from~the perspective of the low-energy phenomenology of hadron physics,
there is an even more important property of QCD. It is the spontaneous 
chiral symmetry breaking of the QCD vacuum and the associated appearance 
of the Nambu--Goldstone bosons. The~chiral quark soliton model (CQSM) is an 
effective theory of QCD, which efficiently incorporates this important dynamical 
symmetry of QCD into the physics of baryon structures~\cite{DPP1988}. 
(Earlier reviews of the model can, for~example, be found in~\cite{CBKPWMAG1996, ARW1996}; see also~\cite{Weigel_Book}.) 
In short, the~theoretical framework of the CQSM is a relativistic mean field theory for 
quark fields. The~quarks in the nucleon or any baryons are supposed to move 
in a mean field of hedgehog shape with nontrivial topology. 
(Remember that the classical pion field configuration of the hedgehog shape
is the core of the Skyrme model as an effective meson theory of baryons~\cite{Witten1983}.)
By taking account of nonperturbative deformation (or vacuum polarization) 
of the negative-energy Dirac-sea quark orbitals under the influence of the 
hedgehog mean field, the~CQSM automatically and effectively incorporates 
pionic quark--antiquark excitation modes inside the baryons. In~fact, the~model 
turned out to reproduce a lot of baryon observables fairly well with almost no 
free~parameters. 

Probably, the~most successful application of the CQSM lies in excellent reproduction 
of the parton distribution functions (PDFs) in the nucleon~\cite{DPPPW1996, DPPPW1997, 
WGR1997L, WGR1997, WK1998, WK1999, Waka2003A, Waka2003B}. 
Among others, predicted flavor asymmetries of the sea-quark (or anti-quark)
distributions for both the unpolarized PDFs and the longitudinally 
polarized PDFs are surprisingly consistent with the observational data accumulated 
to date, which in turn proves the importance of the chiral symmetry embedded 
in the physics of nucleon parton distribution functions~\cite{Waka2003A, Waka2003B}.
In the present paper, we shall review how and why the CQSM can
explain many characteristic features of the parton distribution functions
of the nucleon, mainly focusing on our own~contributions.  

The paper is organized as follows. First, in Section~\ref{sec:CQSM}, we briefly
explain what the CQSM is like. Also shown are several noteworthy features
of the theoretical predictions of the CQSM for nucleon observables.
In Section~\ref{sec:PDF}, we explain how we can calculate the quark
and anti-quark distribution functions in the nucleon. 
The distribution functions given at the low-energy model scale,
which is thought to be around $600 \,\mbox{\rm MeV}$, are then evolved to  
high energy scale, where the experimental data obtained from the deep-inelastic 
scattering measurements exist.
Next, in Section~\ref{sec:SU(3)_CQSM}, the~flavor SU(3) extension of the CQSM is
explained. The~greatest advantage of this extension is that it enables us to make
nontrivial predictions about the asymmetry of the strange and anti-strange quark distributions 
in the nucleon.  
In Section~\ref{sec:GI_decomposition},  short remarks are made on our current understanding
of the gauge-invariant decomposition problem of the nucleon spin.
Next, in~Section~\ref{sec:GPD}, we discuss the generalized form factors 
and Ji's angular momentum sum rule of the nucleon~\cite{Ji1997, Brodsky2010}. 
In Section~\ref{sec:Nspin}, based on Ji's sum rule explained in the previous section, we shall carry 
out a semi-empirical analysis of the nucleon spin contents, especially by paying attention
to their scale dependencies. 
Finally, in~Section~\ref{sec:conclusion}, we make some comments on the future prospects,
based on the discussion in the present~paper.

\section{Brief Introduction to Chiral Quark Soliton~Model}
\label{sec:CQSM}

The chiral quark soliton model (CQSM) is a low-energy effective model of baryons first
proposed by Diakonov, Petrov, and~Pobilitsa based on the instanton-liquid picture of the 
QCD vacuum~\cite{DPP1988}. 
The basic Lagrangian of the CQSM is very simple and given as 
\begin{equation}
 {\cal L}_{\rm CQSM} \ = \ \bar{\psi} (x) \,\left( i \,\slashed{\partial} \ - \ 
 M  \,U^{\gamma_5} (x) \right) \,\psi (x),
\end{equation}
with $\mathnormal{U}^{\gamma_5} (x) \equiv e^{\,i \,\gamma_5 \,\pi (x) / f_\pi}$.
Here, $\pi (x) = \sum_{a = 1,2,3} \mbox{\boldmath $\pi$}^a (x) \,\mbox{\boldmath $\tau$}^a$ 
stands for the isospin-triplet pion fields. (In the flavor SU(3) version of the CQSM, 
the pion fields here should be replaced by the octet meson fields.)
The Lagrangian above describes the effective quark fields $\psi (x)$ nonlinearly coupled to 
the Nambu--Goldsto pion fields $\pi (x)$. 
Here is one important point to note. 
The pions in this model Lagrangian are not independent fields
of quarks, as~indicated by the fact that there is no kinetic
term of the pion fields in the above Lagrangian.
The kinetic terms of the pions are thought to be generated as a quantum effect.
In fact, if~one constructs an effective meson action from the flavor 
SU(3) version of the above effective quark Lagrangian by integrating
out the quark fields with the use of the derivative expansion method, 
it is known to reproduce the famous Skyrmion action with the Wess--Zumino 
term, but~together with soliton destabilizing 4th 
 derivative terms~\cite{Witten1983}. 
Actually, we do not use such an approximate bosonization~formalism.

\begin{figure}[htpb]
\begin{centering}
\includegraphics[width=6.0cm]{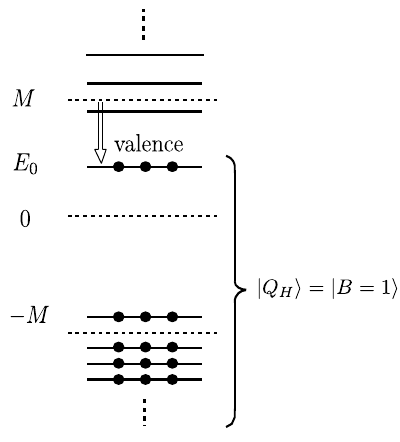}
\caption{Characteristic behavior of the single-quark energy levels under
the hedgehog mean field.\label{fig01}}
\end{centering}
\end{figure}

\vspace{1mm}
The central idea of the CQSM is a soliton construction without
recourse to approximate a bosonization procedure. We start with a classical
pion field configuration with a hedgehog shape as follows, which plays the role of
a mean field for quarks:  
\begin{equation}
 \bm{\pi} (\bm{x}) \ = \ \hat{\bm{r}} \,F(r) .
\end{equation}
Here, 
the~profile function $F (r)$ is supposed to satisfy the following boundary condition,
\begin{equation}
 F (0) \ - \ F (\infty) \ = \ n \,\pi,
\end{equation}
with $n \,( \ = 1)$ being the so-called winding number of the effective pion 
field configuration.
Under the presence of this unique shape of mean field, the~quark field obeys the following
Dirac equation,
\begin{equation}
 H \,\vert m \rangle \ = \ E_m \,\vert m \rangle ,
\end{equation}
with 
\begin{equation}
 H \ = \ \frac{\bm{\alpha} \cdot \nabla}{i} \ + \ M \,\beta \,
 \left( \cos F(r) \ + \ i \,\gamma_5 \,
 \mbox{\boldmath $\tau$} \cdot \hat{\mbox{\boldmath $r$}} \,
 \sin F(r) ) \right) .
\end{equation}
Here, $M$ is supposed to stand for the dynamical quark mass (or the constituent quark mass)
generated by the spontaneous chiral symmetry breaking of the QCD vacuum.
A characteristic feature of the above Dirac equation is that one deep bound state 
emerges from the positive-energy Dirac continuum (see Figure~\ref{fig01}). 
Hereafter, we call this quark 
orbit the valence quark level. Then, a~baryon-number-one object with respect to
the physical vacuum is obtained by putting $N_c (=3)$-quarks into this
valence orbital as well as all the negative-energy Dirac-sea~orbitals.

\vspace{2mm}
Accordingly, the~total energy of this baryon-number-one
object is given as a sum of the energy of $N_c$ valence quarks
and the energy of deformed (vacuum-polarized) Dirac-sea quarks as
\begin{equation}
 E_{static} \ = \ N_c \,E_0 \ + \ E_{v.p.}.
\end{equation}
Here, $E_0$ represents the single-particle energy of the valence quark orbital,
while the vacuum polarization contribution corresponds to the
Casimir energy resulting from the polarization (deformation)
of the Dirac-sea quark orbitals. The~latter is given as
\vspace{+6pt}
\begin{equation}
 E_{v.p.} \ = \ N_c \,\left( \sum_{m \,(E_m < 0)} \,E_m \ - \ 
 \sum_{k \,(\epsilon_k < 0)} \epsilon_k \right). \label{Eq:E_vpl}
\end{equation}
That is, the~Casimir energy is given as a sum of all the energies
of quarks in the negative-energy Dirac-sea orbitals. 
The 2nd term here represents the subtraction of 
the Dirac-sea energy of the physical vacuum, which is obtained by 
letting $F (r) \rightarrow 0$.
The most probable pion field configuration is then determined from
the stationary requirement for the total energy functional $E_{static} [F(r)]$,
\begin{equation}
 \frac{\delta}{\delta F(r)} \,E_{static} [ F (r)] \ = \ 0.
\end{equation}
This requirement combined with the above Dirac
equation is reduced to a self-consistent Hartree problem
which can be solved by the numerical method of Kahana, Ripka, and~Soni~\cite{KR1984, KRS1984}.
(See~\cite{WY1991} for more details about the actual calculation method.) 
After self-consistency is fulfilled, the~hedgehog pion field,
which was originally introduced as an external mean field for
quarks, becomes an implicit functional of the quark~fields.

Actually, the~vacuum polarization energy given by Equation~(\ref{Eq:E_vpl}) contains
ultraviolet divergence (logarithmic
divergence).
Often, this ultraviolet divergence is removed with the use of the Pauli--Villars 
regularization, which means the following replacement of  the effective 
action~\cite{DPPPW1996, DPPPW1997} 
\begin{equation}
 S_{eff} [\bm{\pi}] \ \rightarrow \ S^M_{eff} [\bm{\pi}] \ - \ 
 \left( \frac{M}{M_{PV}} \right)^2 \,
 S^{M_{PV}}_{eff}  [\bm{\pi}] ,
\end{equation}
where $M_{PV}$ is a Pauli--Villars cutoff mass. For~a given dynamical quark mass $M$,
the Pauli--Villars mass $M_{PV}$ is fixed from the following condition
\begin{equation}
 \frac{N_c}{4 \,\pi^2} \,M^2 \,\log \frac{M^2_{PV}}{M^2} = f^2_\pi.
\end{equation}
with $f_\pi$ being the pion weak-decay constant. This condition follows from
the requirement that the above regularized action reproduces the correct
pion kinetic term after bosonization. 
For some special quantities, like the vacuum quark condensate as well as the 
nucleon scalar charge, however, the~above single-term Pauli--Villars subtraction 
is not enough, because~those quantities contain quadratic divergence.
For handling those special quantities, we must use more sophisticated 
double-term Pauli--Villars subtraction, as discussed in~\cite{KWW1999}, which
requires the following replacement of the effective action:   
\begin{equation}
 S_{eff} [\bm{\pi}] \ \rightarrow \ S^M_{eff} [\bm{\pi}] \ - \ 
 \sum_{i = 1}^2 \,c_i \, S^{\Lambda_i}_{eff}  [\bm{\pi}] ,
\end{equation}
Four subtraction parameters $c_1, c_2, \Lambda_1, \Lambda_2$  are determined 
so as to remove quadratic and logarithmic divergence of the effective action and 
to reproduce the empirical value of vacuum condensate and correct pion kinetic
energy term in the effective pion action~\cite{KWW1999}. 
Once these parameters are fixed, the~model does not contain any other free~parameters.

\vspace{2mm}
The quark hedgehog state $\vert Q_H \rangle \equiv \vert B \!=\!1 \rangle$
constructed on the basis of the hedgehog mean field breaking 
 the rotational
symmetry in the coordinate space as well as in the isospin space, so that
it is not a good spin--isospin eigen-state. This comes from the degeneracy of
the static energy under the rotation in the coordinate and isospin spaces. 
(Remember the analogous situation that happens for the baryon-number-one object 
in the Skyrme model
~\mbox{\cite{ANW1983, AN1984}}.)
This naturally generates a spontaneous (zero-mode) rotation of the hedgehog mean field,
which can be parametrized as
\begin{equation}
 U^{\gamma_5} (\bm{x}, t) \ = \ A(t) \,U^{\gamma_5}_0 (\bm{x}) \,A^\dagger (t)
 \ \ \ : \ \ \ A (t) \in SU(2) ,
\end{equation}
where $A (t)$ is a time-dependent $SU(2)$ matrix describing the rotation of
the hedgehog mean field in the coordinate and isospin spaces.
Now the spin--isospin projection of the rotating hedgehog is carried out by using the cranking
method familiar in nuclear physics~\cite{RS_Book}, which consists of the following procedures:  

\begin{itemize}
\item[$\bullet$] Cranked iso-rotation of hedgehog mean field induces
Coriolis coupling acting on the quarks in the rotating frame given by
\vspace{-3pt}%
\begin{equation}
 \Omega \ \equiv \ - \,i \, A^\dagger (t) \,\dot{A} (t) \ \equiv \ 
 \frac{1}{2} \,\Omega_a \,\tau_a.\vspace{-3pt}
\end{equation}
\item[$\bullet$] Evaluate changes of the intrinsic quark wave function and the
associate changes of observables by treating the above Coriolis coupling as
an external perturbation.
\item[$\bullet$] Canonically quantize the iso-rotational motion. 
\end{itemize} 

Although the detail is skipped here, the~final formula for evaluating baryon observables
is given in the following form, i.e.,~in the form that the effective operator 
$\langle O \rangle_A$ as a function of the collective coordinates $A$ is sandwiched by 
the wave functions describing the collective rotational motion of hedgehog mean field 
as~\cite{DPP1988, WY1991}
\begin{equation}
 \langle J^\prime \,M^\prime_J \,M^\prime_T \,\vert \,O \,\vert J \,M_J \,M_T \rangle
 \ = \ \int {\cal D} A \,\,\Psi^{(J^\prime)*}_{M^\prime_J \,M^\prime_T} [A] \,
 \langle O \rangle_A \,\Psi^{(J)}_{M_J \,M_T} [A].
\end{equation}
Here, $\Psi^{(J)}_{M_J \,M_T} [A]$ is the wave function describing the collective 
rotational motion of the baryon states and is given as
\begin{equation}
 \Psi^{(J)}_{M_J \,M_T} [A] \ = \ \sqrt{\frac{2 \,J + 1}{8 \,\pi^2}} \,\,(- \,1)^{T + T_3} \,\, 
 D^{(J)}_{- \,T_3 \,J_3} [A] ,
\end{equation}
where $D^{(J)}_{- \,T_3 J_3} [A]$ is the familiar Wigner rotation matrix.
The effective operator $\langle O \rangle_A$ consists of the zeroth and the 
first-order terms in the collective angular velocity $\Omega$ as
\begin{equation}
 \langle O \rangle_A \ = \ \langle O \rangle^{(0)}_A \ + \ \langle O \rangle^{(1)}_A
 \ + \ \cdots.
\end{equation}
The lowest order term $\langle O \rangle^{(0)}_A$ just corresponds to the 
answer in the mean field theory,
and it is given as a diagonal sum over the occupied states consisting of 
the valence quark orbital $\vert 0 \rangle$ with the energy $E_0$
and all the negative-energy Dirac-sea orbitals $\vert n \rangle$ with $E_n < 0$ as
\begin{equation}
 \langle O \rangle^{(0)}_A \ = \ N_c \, \sum_{n \,\in \,occupied} \,
 \langle n \,\vert \,\tilde{O} \,\vert \,n \rangle,
\end{equation}
with
\begin{equation}
 \tilde{O} \ \equiv \ A^\dagger \gamma^0 \,O \,A.
\end{equation}

In the above equation, $\sum_{\,n \in occupied}$ stands for the sum over the 
occupied quark orbitals $n$.
On the other hand, the $O (\Omega^1)$ term or the $1 / N_c$ correction term is given as
a double sum over the occupied orbitals $n$ and the non-occupied orbitals $m$
as follows: 

\begin{equation}
 \langle O \rangle^{(1)}_A \ = \ \frac{N_c}{2} \,
  \sum_{\substack{m \,\in \,non-occupied, \\ \,n \,\in occupied}} \ 
 \frac{1}{E_m - E_n} \,\left[ \langle n \,\vert \,\tilde{O} \,\vert m \rangle \,
 \langle m \,\vert \,\Omega \,\vert \,n \rangle \ + \ 
 \left( \tilde{O} \, \leftrightarrow \, \Omega \,\right) \right] ,
\end{equation}
which describes virtual transitions from the occupied states to the
non-occupied states by the action of the external field and the Coriolis force
and vice~versa.

\vspace{1mm}
Here we summarize several noteworthy features of the predictions of the CQSM
for nucleon~observables.

\begin{itemize}
\item[(1)] Good reproduction of the neutron charge distribution
as a clear evidence of efficiently incorporating the pion cloud effect.
\item[(2)] Resolution of the famous underestimation problem of the isovector 
axial-vector coupling constant $g^{(3)}_A$ of the nucleon in the Skyrme model.
\item[(3)] Reproduce a large $\pi N$ sigma term consistent with the
empirical information as well as highly nontrivial behavior of the scalar quark
density of the nucleon in coordinate space and momentum space.  
\item[(4)] Good reproduction of the small quark spin fraction of the nucleon
consistent with the high-energy deep-inelastic-scattering data by the EMC group. 
\end{itemize}

\noindent
Following are supplementary explanations concerning the 
above-mentioned remarkable features of the model predictions.

\begin{itemize}
\item[(1)] 
We first explain the reason why good reproduction of the neutron charge 
distribution is a noteworthy matter~\cite{Waka1991}. 
Shown in Figure~\ref{fig02} is the prediction of the CQSM for the neutron
charge distribution. The~dashed curve represents the contribution of
the $N_c \,( = 3)$ valence quarks to the neutron charge density.
Note that the contribution
from the valence quarks is positive in most spatial regions and its magnitude 
rapidly decreases as the distance $r$ from the neutron center becomes large. 
The dash-dotted curve in the same figure shows
the contribution from the negative-energy Dirac-sea quarks.
As one sees, this contribution is negative in all the region but it has a
long-range tail as $r$ becomes large as compared with the contribution of the
valence quarks.

Probably, the~above feature can be understood based on
the well-known meson theory of Yukawa. In~this theory, the~physical neutron is
thought to virtually dissociate into the superposition state of the proton and 
the $\pi^-$ as $n \ \rightarrow \ p \ + \ \pi^-$. 
Since the $\pi^-$ is much lighter than the proton (and the neutron),  
it virtually travels far away from the center of the neutron.
This results in a centrally concentrated positive charge distribution by the 
virtual proton and a negative charge distribution by the virtual 
$\pi^-$ that dominates in the outer region. (The charge conservation naturally
ensures that the net charge of the neutron is zero.) 
We think that the contribution of the three valence quarks in the CQSM simulates
the positive charge distribution due to the virtual proton, while the contribution from the
Dirac-sea quark simulates the negative charge distribution due to the cloud of
$\pi^-$. Undoubtedly, the~neutron charge distribution is thought to provide the 
simplest clear evidence showing the importance of the chiral symmetry of QCD, 
which is efficiently taken into account into the framework of the~CQSM. 

\vspace{-3pt}
\begin{figure}[htbp]
\includegraphics[width=9.0cm]{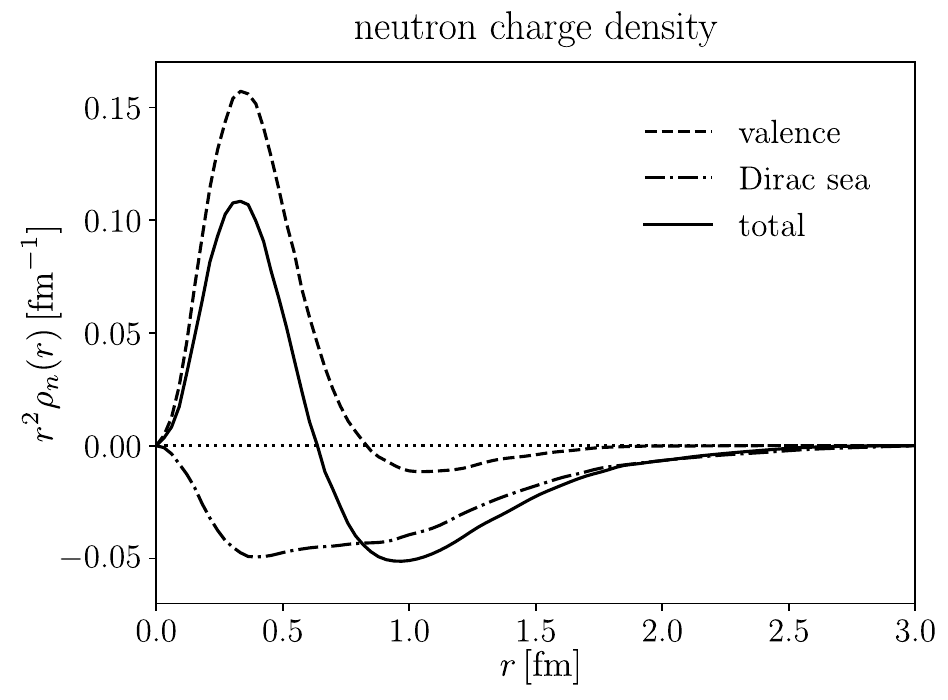}
\caption{The CQSM prediction for the neutron charge density
$\rho_n (r)$ multiplied by $r^2$. The~dashed and dash-dotted curves respectively 
stand for the contribution of the $N_c$ valence quarks and
the negative-energy Dirac-sea quarks, while the solid curve represents their sum.\label{fig02}}
\end{figure}

\item[(2)]
Although it was not necessarily taken seriously, there was a sizable 
underestimation problem of the isovector axial-vector coupling constant 
$g^{(3)}_A$ of the  nucleon in the Skyrme model~\cite{ANW1983,AN1984}. 
As compared with the empirically known value
$g^{(3)}_A = 1.27$, the~prediction of the Skyrme model is known to be
around $0.6 \sim 0.8$.
One might think that this is a minor problem concerning a tiny flaw of a model.
However, this problem was not solved within the Skyrme model, and~it turned out
to persist in any soliton models based on the hedgehog configuration within 
effective Lagrangians of mesons~\cite{SS2005A, SS2005B, HSS2008}.
As already pointed out, there are many common features between the Skyrme
model and the CQSM. Somewhat unexpectedly, however, we noticed that, within~
the framework of the CQSM,  there is an important 1st-order rotational
correction in the collective angular velocity $\Omega$ (it is also
thought of as a novel $1 / N_c$ correction), which is completely missing in the
framework of the Skyrme model~\cite{WW1993, CBGPPWW1994}. 
In fact, it turned out that these two 
models give the following prediction on $g^{(3)}_A$:  
\begin{eqnarray}
 g^{(3)}_A ({\rm Skyrme}) &=& g^{(3)}_A (\Omega^0) \ + \ g^{(3)}_A (\Omega^1)
 \ \simeq \ (0.6 \, \sim \, 0.8) \ + \ \ 0 \ \simeq \ 0.6 \, \sim \, 0.8, \\
 g^{(3)}_A ({\rm CQSM}) &=& g^{(3)}_A (\Omega^0) \ + \ g^{(3)}_A (\Omega^1)
 \ \simeq \ 0.8 \ + \ 0.4 \ \simeq \ 1.2.
\end{eqnarray}
It was argued that there is a deep reason
for this critical difference between the prediction of the CQSM as an effective fermion 
theory and that of the Skyrme model as an effective meson theory~\cite{Waka1995, Waka1996}.
From more fundamental standpoint based on an effective theory at the quark level,
we can say that the ultimate origin of the $g_A$ problem in the
Skyrme model comes from the non-commutativity of the bosonization
procedure and the collective quantization procedure. (For more detail, we refer to
the literature above~\cite{Waka1995, Waka1996}.)
In short, an~important piece of information 
 of the original fermion theory is
lost in the process of the bosonization procedure. Aside from this fundamental
difference, we emphasize that the CQSM generally gives more realistic physical 
predictions on most baryon observables than the Skyrme model. 
Besides, as~we shall discuss in the following,
the most important advantage of the CQSM as compared with the Skyrme model
is that the former can handle the non-local quark--quark correlation inside
the nucleon, which is necessary to evaluate the quark distribution functions of 
the nucleon, while there is no way to handle such non-local quark--quark correlation 
within the theoretical framework of effective meson theories of baryons, including 
the Skyrme~model.

\item[(3)] 
Probably, a~highly unique feature of the CQSM is that it simultaneously reproduces
nontrivial local chiral structure of the nucleon and the QCD vacuum structure with
nonzero quark condensate~\cite{KWW1999, RW2012}. 
It can be seen from Figure~\ref{fig03}, which shows the model prediction 
of the nucleon scalar charge density in the QCD vacuum.
The dashed curve and the~dash-dotted curve 
 here respectively denote the contribution of the $N_c$ 
valence quarks and that of the Dirac-sea quarks to the nucleon scalar quark density,
while their sum is shown by the solid curves. One can see that the contribution of the
valence quarks smoothly attenuates to zero as the distance from the nucleon center
becomes large, as is the case with the predictions of almost all models of
the nucleon. Remarkably, however, the~contribution of the negative-energy Dirac-sea
quarks does not attenuate to zero, but~it rather approaches a negative value, 
 which is nothing but the value of the QCD vacuum quark
condensate. This means that the CQSM can explain the vacuum quark condensate and
the nontrivial local structure of the nucleon scalar charge density simultaneously.
One may naturally anticipate that this extraordinary structure of the nucleon scalar charge
density would show itself in some observables. It was shown to appear as a delta-function
type singularity in the twist-3 chiral-odd quark distribution function of the 
nucleon~\cite{Schweitzer2003, WO2003, OW2004}.
Since this topic was intensively discussed in a recent review paper~\cite{Waka2024}, 
we do not discuss it further in the present~paper.

\vspace{-3pt}
\begin{figure}[htbp]
\includegraphics[width=8.0cm]{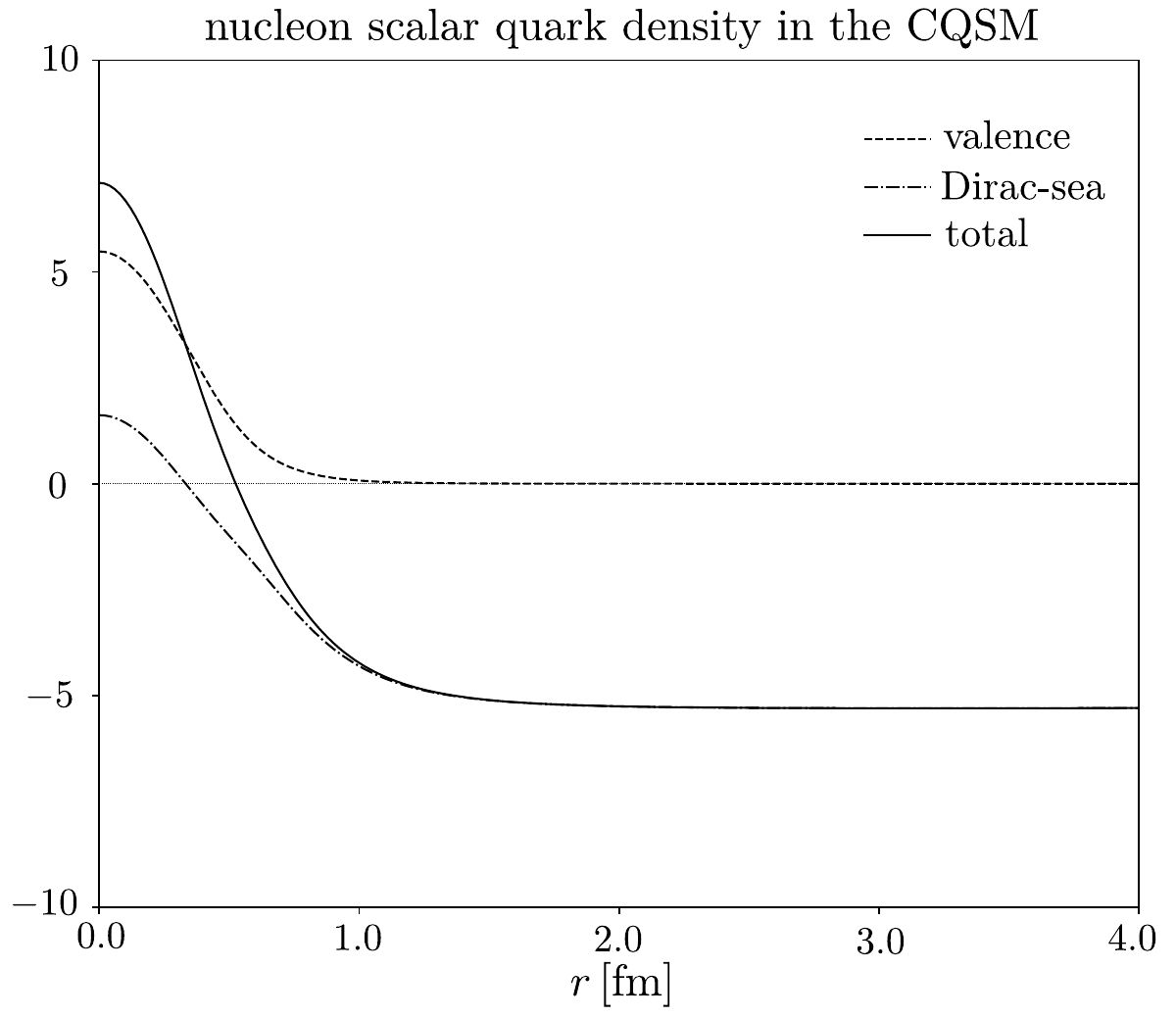}
\caption{The CQSM prediction of the nucleon scalar charge density.
The dashed and dash-dotted curves respectively stand for the contributions from
the $N_c$ valence quarks and the Dirac-sea quarks, while the solid curve 
represents the total contribution. \label{fig03}}
\end{figure}  

\item[(4)] 
Another prominent feature of the CQSM is that it predicts fairly small quark
spin fraction of the nucleon as~\cite{WK1999}
\begin{equation}
 \Delta \Sigma \ \simeq \ 0.35,
\end{equation}
at the energy scale of the model, which is thought to be around $600 \,\mbox{\rm MeV}$.
Since the CQSM is an effective theory of quarks, which does not explicitly contain the
gluon degrees of freedom, 
 it satisfies the following sum rule of the nucleon spin~\cite{WY1991}
\vspace{+6pt}
\begin{equation}
 \frac{1}{2} \,\Delta \Sigma \ + \ L^Q \ = \ \frac{1}{2},
\end{equation}
where $\frac{1}{2} \,\Delta \Sigma$ represents the contribution of the intrinsic quark spin, 
while $L^Q$ represents the contribution of the quark orbital angular momentum to the
net nucleon spin.
The smallness of the quark spin fraction therefore implies the largeness of the contribution
of the orbital angular momentum of quarks.
Undoubtedly, the~largeness of the orbital angular momentum contribution is inseparably 
connected with the basic dynamical assumption of the CQSM, i.e.,~its nucleon picture as 
a rotating hedgehog~object.
\end{itemize}

So far, we have demonstrated that the CQSM model is able to explain important
characteristics of several nucleon observables, in~which the dynamical chiral symmetry
of QCD plays a critically important role. Still, most low-energy baryon observables are
insensitive to the differences between low-energy effective models like the MIT bag
model and the non-relativistic quark model, etc.
In the following sections, we shall show that the potential ability of the CQSM manifests
most clearly in its predictions about the internal partonic structure of the~nucleon.

\section{CQSM and Nucleon Parton Distribution~Functions}
\label{sec:PDF}

For obtaining the distribution functions of quarks, we need to evaluate nucleon matrix 
elements of quark bilinear operators $\psi^\dagger (0) \,O \,\psi (z)$ with the light-cone 
separation given as
\begin{equation}
 q (x) \ = \ \frac{1}{4 \,\pi} \,\int_{- \,\infty}^\infty \,d z^0 \,e^{\,i \,x \,M_N \,z^0} \,
 \langle N (P) \,\vert \psi^\dagger (0) \,O \,\psi (z) \,\vert N (P) \rangle
 \vert_{z^3 = - \,z^0, \, z_\perp = 0} \, ,
\end{equation}
with the definition of the standard light-cone coordinates $z^{\pm} =(z^0 \pm z^3) / \sqrt{2}$. 
Here, $\vert N(P) \rangle$ represents the nucleon state with momentum $P$. 
(It is the Lorentz-boost invariance of the quark distribution function along 
the $z^3$-axis, i.e.,~the direction of the momentum of the parent nucleon,
that allows us to evaluate the above matrix elements in the nucleon rest frame.)
We set $O = \gamma^+$ and $O = \gamma^+ \,\tau_3$ for the isoscalar unpolarized
distribution function $u (x) + d (x)$ and the isovector unpolarized distribution function
$u (x) - d (x)$. On~the other hand, we take $O = \gamma^+ \,\gamma_5$ and
$O = \gamma^+ \,\gamma_5 \,\tau_3$ for the isoscalar longitudinally polarized distribution
function $\Delta u (x) + \Delta d (x)$ and the isovector longitudinally polarized distribution
function $\Delta u (x) - \Delta d (x)$.
On account of the charge conjugation properties of relevant operators, we can formally
extend the defining region of quark distribution functions to the interval
$- \,1 \leq x \leq 1$ as
\begin{eqnarray}
 \bar{u} (x) + \bar{d} (x) \ &=& \ - \,\left[ u (- \,x) \ + \ d (- \,x) \right], \\
 \bar{u} (x) - \bar{d} (x) \ &=& \ - \,\left[ u (- \,x) \ - \ d (- \,x) \right], \\
 \Delta \bar{u} (x) + \Delta \bar{d} (x) \ &=& \ \Delta u (- \,x) \ + \ \Delta d (- \,x), \\
 \Delta \bar{u} (x) - \Delta \bar{d} (x) \ &=& \ \Delta u (- \,x) \ - \ \Delta d (- \,x).
\end{eqnarray}
(For readers who are not familiar with the above relations, we shall explain its theoretical 
basis in Appendix \ref{appA}.)

In the above four equations, the~variable $x$ is supposed to lie in the physical range 
$0 \leq x \leq 1$. These relations mean that the quark distributions
in the negative $x$ region can actually be interpreted as the corresponding anti-quark 
distributions after taking care of differences in signs.
We also point out that the following novel $\Omega$ ($\sim$$1/N_c$) dependencies follow
from the theoretical structure of the model, i.e.,~the mean field of hedgehog shape and the
subsequent perturbative treatment of the collective rotational 
motion~\cite{DPPPW1996, DPPPW1997}:  
\vspace{+6pt}
\begin{eqnarray}
 u (x) \ + \ d (x) \ &\sim& \ O (\Omega^0) \ + \ 0, \\
 u (x) \ - \ d (x) \ &\sim& \ \ 0 \ + \ O (\Omega^1), \\
 \Delta u (x) \ + \ \Delta d (x) \ &\sim& \ \ 0 \ + \ O (\Omega^1), \\
 \Delta u (x) \ - \ \Delta d (x) \ &\sim& \ O (\Omega^0) \ + \ O (\Omega^1).
\end{eqnarray}
Here, $u (x)$ and $d (x)$ respectively stand for the unpolarized distribution functions of
the $u$-quark and the $d$-quark, while $\Delta u (x)$ and $\Delta d (x)$ represent
the longitudinally polarized distribution functions of the $u$-quark and the $d$-quark.

Just for reference, we write down the theoretical expressions for the above four
basic distribution functions of the nucleon. They are given as~\cite{DPPPW1996, DPPPW1997,
WGR1997L, WGR1997, WK1998, WK1999}
\begin{eqnarray}
 u (x) + d (x) &=& M_N \,N_c \,\sum_{n \in occupied} \,
 \langle n \,\vert \,(1 + \gamma^0 \,\gamma^3) \,\delta_n \,\vert n \rangle, \\
 u (x) - d (x) &=& M_N \,\frac{1}{I} \,\sum_{a = 1}^3 \frac{N_c}{2} 
 \sum_{\substack{m \in non-occupied \\ n \in occupied}}
 \langle n \,\vert \,\tau_a \,(1 + \gamma^0 \,\gamma^3) \,
 \frac{\delta_n + \delta_m}{2} \,\vert m \rangle \,
 \langle m \,\vert \,\tau_a \,\vert \,n \rangle, \ \ \ \ \ \ \ \ \ \\
 \Delta u (x) + \Delta d (x) &=& - \,M_N \,\frac{1}{I} \,\frac{N_c}{2}
 \sum_{\substack{m \in non-occupied \\ n \in occpied}}
 \langle n \,\vert \,(1 + \gamma^0 \,\gamma^3) \,
 \frac{\delta_n + \delta_m}{2} \,\vert m \rangle \,
 \langle m \,\vert \,\tau_3 \,\vert \,n \rangle, \\
 \Delta u (x) - \Delta (x) &=& \frac{1}{3} \,M_N \,N_c \,
 \sum_{n \in occupied} \,
 \langle n \,\vert \,\tau_3 \,(1 + \gamma^2 \,\gamma^3) \,
 \gamma_5 \,\delta_n \,\vert n \rangle \ + \ O(\Omega^1), 
\end{eqnarray}
with the definition $\delta_n = \delta (x \,M_N - E_n - p_3)$. 
Since the expression of the $O (\Omega^1)$ contribution to $\Delta u(x) - \Delta d (x)$
is fairly complicated, it is omitted~here.

\begin{figure}[htbp]
\includegraphics[width=12.0cm]{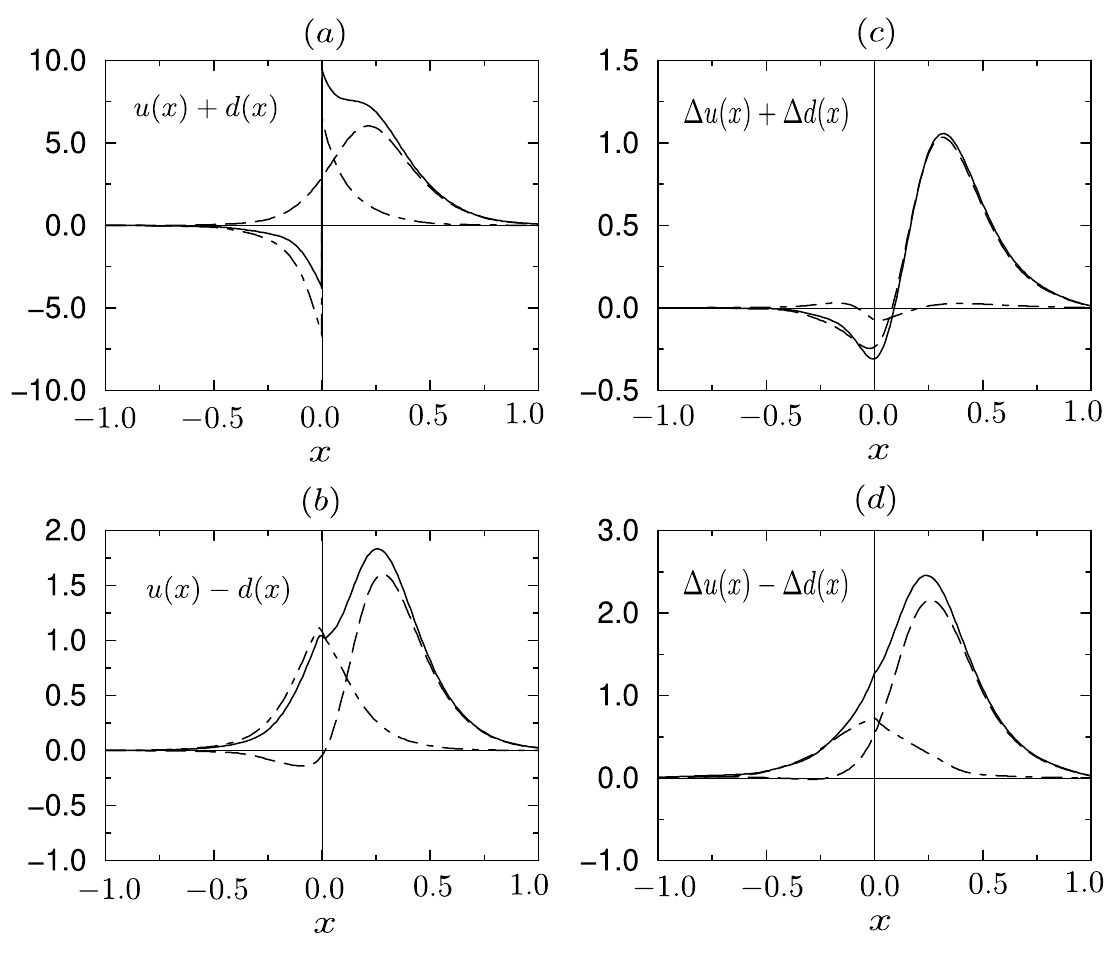}
\caption{The CQSM prediction for the twist-2 quark distribution functions of the nucleon.
(\textbf{a})~Isoscalar unpolarized distribution, (\textbf{b}) isovector unpolarized distribution,
(\textbf{c}) isoscalar longitudinally polarized distribution, (\textbf{d}) isovector longitudinally polarized
distribution. In~these figures, the~dashed curve, the~dash-dotted, and~the solid curves
respectively represent the contributions from the valence quarks, the~Dirac-sea quarks,
and the sum of them.\label{fig04}}
\end{figure}

\vspace{2mm}
Shown in Figure~\ref{fig04} are the CQSM predictions for the basic twist-2 quark
distribution function of the nucleon. The~four panels (a), (b), (c), and (d) 
 respectively
stand for the isoscalar unpolarized distribution $u (x) + d (x)$, the~isovector
unpolarized distribution $u(x) - d (x)$, the~isoscalar longitudinally polarized
distribution $\Delta u (x) + \Delta d (x)$, and~the isovector longitudinally
polarized distributions $\Delta u(x) - \Delta d (x)$.
These figures already show highly nontrivial structure of the quark distribution
functions predicted by the CQSM in the small $x$ region and in the negative $x$ region. 
Remember that the quark distributions
in the negative $x$ region can be interpreted as anti-quark distributions
aside from signs. In~particular, these nontrivial predictions for the anti-quark
distributions are inseparably connected with the basic feature of the CQSM,
which enables us to take account of nonperturbative vacuum-polarization of
the negative-energy Dirac-sea quarks in the hedgehog mean~field.    

\begin{figure}[htbp]
\includegraphics[width=8.5cm]{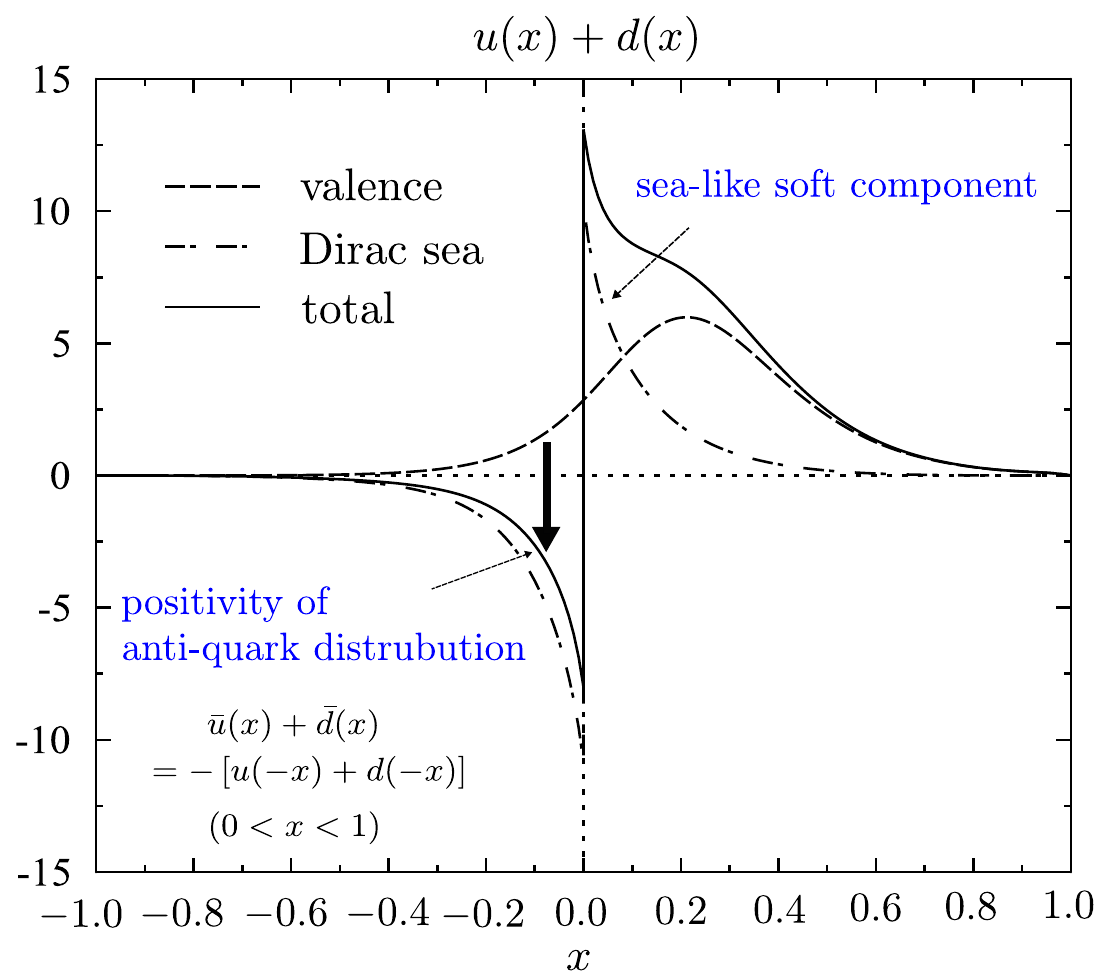}
\caption{On the noteworthy feature of the CQSM prediction for the distribution 
$u (x) + d (x)$, which ensures the positivity of the anti-quark distribution
functions $\bar{u} (x) + \bar{d} (x)$ with a physical value of $x$.\label{fig05}}
\end{figure}

\vspace{2mm}
One can see that, in~any of these four distribution functions, the~Dirac-sea quarks
give important and characteristic contributions.
Worthy of special mention at this stage is the isoscalar unpolarized distribution 
$u (x) + d (x)$.
The contribution of the valence quarks to this distribution has a peak around
the value of $x \sim 0.25$, but~it has a tail with a positive sign extending 
to the negative region of $x$. We emphasize that basically the same behavior is also 
predicted by most three-quark models of the nucleon, like the 
non-relativistic quark model and also that of the MIT bag model.
However, these predictions are unacceptable, if~we remember the relation
$\bar{u} (x) + \bar{d} (x) = - \,[ u (-x) + d (-x)]$ with $0 < x < 1$, which
means that $u (x) + d (x)$ in the negative $x$ region is identified with
the anti-quark distribution $\bar{u} (x) + \bar{d} (x)$ for a physical value of 
$0< x < 1$, but~with an extra minus sign. 
Accordingly, once we discard the contribution of the Dirac-sea
contribution, the~positivity of the valence quark contribution in the negative $x$
region breaks the positivity of the anti-quark distribution for the physical value of $x$. 
Amazingly, however, if~the Dirac-sea contribution
is properly taken into account, the~net contribution to $u (x) + d (x)$ in the negative
$x$ region is definitely negative, which means that the prediction of the CQSM
legitimately satisfies the required positivity requirement for the anti-quark
distribution $\bar{u} (x) + \bar{d} (x)$ (See Fig.\ref{fig05}.). 
This feature is one of the great advantages of the CQSM as a field-theoretical model
of the nucleon~\cite{DPPPW1996}.  

\vspace{-6pt}
\vspace{-9pt}

\vspace{6mm}
Unfortunately, the~above interesting predictions of the CQSM cannot be immediately
compared with the empirically known distribution functions, which are obtained from the
analyses of the high energy deep-inelastic-scattering (DIS) data.
The reason is that the quark distribution functions are in general
renormalization-scale (or energy-scale)-dependent quantities. 
In fact, the~above predictions of the CQSM for the quark distribution functions of the 
nucleon are interpreted to correspond to the distribution functions at the energy scale
around 600 \,MeV, or~$Q^2 \simeq (600 \,\mbox{\rm MeV})^2$, which is to be identified
with the energy scale of the effective model.
On the other hand, the~quark distribution function extracted from the high-energy
deep-inelastic-scattering (DIS) experiments are known to correspond to a high energy scale, 
say at least above $Q^2 \simeq (1 \,\mbox{\rm GeV})^2$. 
A frequently used strategy is to connect the model predictions given at the low energy scale
and the empirically known distribution functions through the 
Dokshitzer–Gribov–Lipatov–Altarelli–Parisi (DGLAP) evolution equation, which is
basically the perturbative renormalization group (RG) equation. 
An immediate question here is whether it is legitimate to use such a perturbative RG
equation at a low energy scale, especially because the QCD running coupling constant
$\alpha_S (Q^2)$ is known to show diverging behavior as $Q^2 \rightarrow 0$ within
the perturbative treatment of~QCD.

\begin{figure}[htbp]
\includegraphics[width=8.0cm]{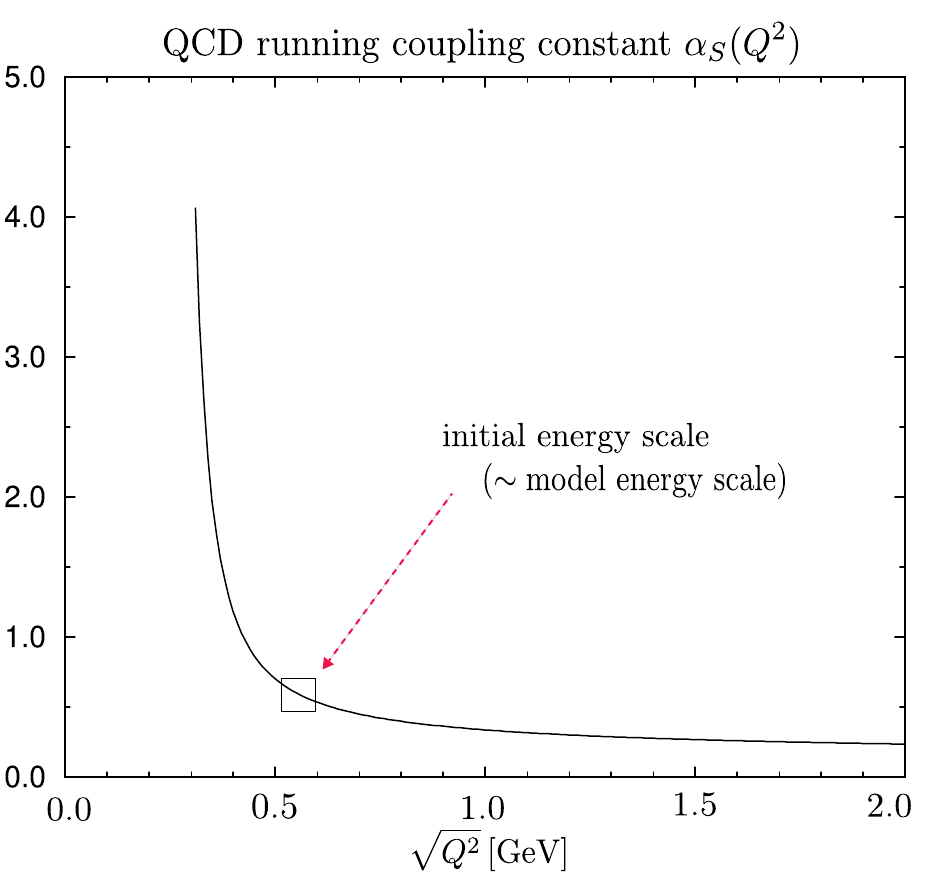}
\caption{The QCD running coupling constant $\alpha_S (Q^2)$ at the next-to-leading
order (NLO) as a function of the renormalization scale $\sqrt{Q^2}$.\label{fig06}}
\end{figure}

\vspace{2mm}
Shown in Figure~\ref{fig06} is the QCD running constant $\alpha_S (Q^2)$ at the
next-to-leading order (NLO) as a function of $\sqrt{Q^2}$. One sees that, at~the low 
energy scale around 600 MeV, the~perturbative QCD treatment looks barely applicable. 
(This should be contrasted with the fact that the initial scale of evolution in most 
low-energy models of QCD like the MIT bag model or the cloudy bag model is around
$400 \,\mbox{\rm MeV}$. The~running coupling constant $\alpha (Q^2)$
at this energy is seen to be close to unity, which cannot be thought of as a small
parameter in perturbation theory.) This would enable us
to use the DGLAP equation at NLO to relate the predictions of the CQSM for PDFs with
the empirical PDFs corresponding to high energy~scales. 

\vspace{-6pt}

The CQSM predictions for the basic twist-2 PDFs are evolved to high energy scales
by using the fortran programs provided in the references~\cite{MK1996, HKM1998}.
A difficult problem here is that, to~solve the evolution equations in the
flavor singlet channel, we also need the gluon distribution functions
at the starting low energy scale, but~a nonperturbative evaluation of the gluon 
distribution is impossible in any low-energy effective model of baryons, 
including the CQSM. The~frequently used strategy is
to assume that the gluon distributions at the low energy scale are
negligibly small and to set zero.  
Based on our experience so far, there is a noteworthy difference between 
the two types of gluon distributions, i.e.,~the unpolarized gluon distribution 
and the longitudinally polarized gluon distribution.
First, consider the net gluon spin $\Delta G$ as the 1st moment or the integral 
of the longitudinally polarized gluon distribution $\Delta g (x)$.
An interesting trial analysis would be as follows. Suppose that, starting from
the empirical information for $\Delta G$ given at the high energy scales, 
for example, the~information given by DSSV fit at $Q^2 = 10 \,\mbox{GeV}$, 
we solved the evolution equation at the next-to-leading order 
to estimate the value of $\Delta G$ at the lower energy scales.
(This means somewhat unconventional downward evolution or disevolution
of the parton distribution functions.)
Then, we observe that, as~$Q^2$ decreases, the~magnitude of $\Delta G$ gradually
reduces, and~it eventually becomes zero approximately around the energy scale 
$Q^2 \simeq (600 \,\mbox{MeV})^2$. This indicates that, just around the model energy
scale of the CQSM, the~contribution of the gluon to the net nucleon spin
is fairly small, which in turn indicates the smallness of the longitudinally
polarized gluon distribution. 
In fact, that this consideration is not off the mark can also be
confirmed by the following observation. In~our treatment, the~predictions of 
the flavor singlet piece of the longitudinally polarized quark distributions 
at a high energy scale are obtained by solving the coupled evolution equations for
the quarks and gluon. The~initial quark distributions at the low energy scale are taken
from the theoretical predictions of the CQSM, while the longitudinally polarized
gluon distribution at this low energy scale is assumed to be zero.
Curiously, we found that such a strategy works fairly well, at least to 
reproduce the empirically known behavior of the longitudinally polarized quark
distributions at the high energy~scales.

The situation is a little different for the unpolarized quark and gluon
distribution functions.
Starting from the empirical information for the quark and gluon momentum
fractions, $\langle x \rangle^Q$ and $\langle x \rangle^G$, suppose that 
we carry out a similar downward evolution to estimate the magnitudes of 
the quark and gluon momentum fractions at the lower energy scales. 
Then, we found that, even at the low energy scale
around $600 \,\mbox{MeV}$, the~gluon still maintains non-negligible momentum 
fraction. This naturally indicates that the unpolarized gluon 
distribution is likely to have a sizable magnitude even at such low energy scales.
This would also make the CQSM prediction for the flavor singlet combination
of the unpolarized quark distributions less reliable as compared with the 
corresponding predictions for the flavor--nonsinglet 
 combination of the unpolarized
quark distributions as well as with the longitudinally polarized quark distributions.  
As is widely believed, nonperturbative evaluation of the gluon distribution is 
feasible only within the framework of lattice QCD. However, it may take some time 
to be able to make really trustworthy predictions, especially for the quark 
and gluon distribution functions in the flavor--singlet channel.  
Keeping these cautions in mind, let us move forward.

\begin{figure}[htbp]

\includegraphics[width=14.5cm]{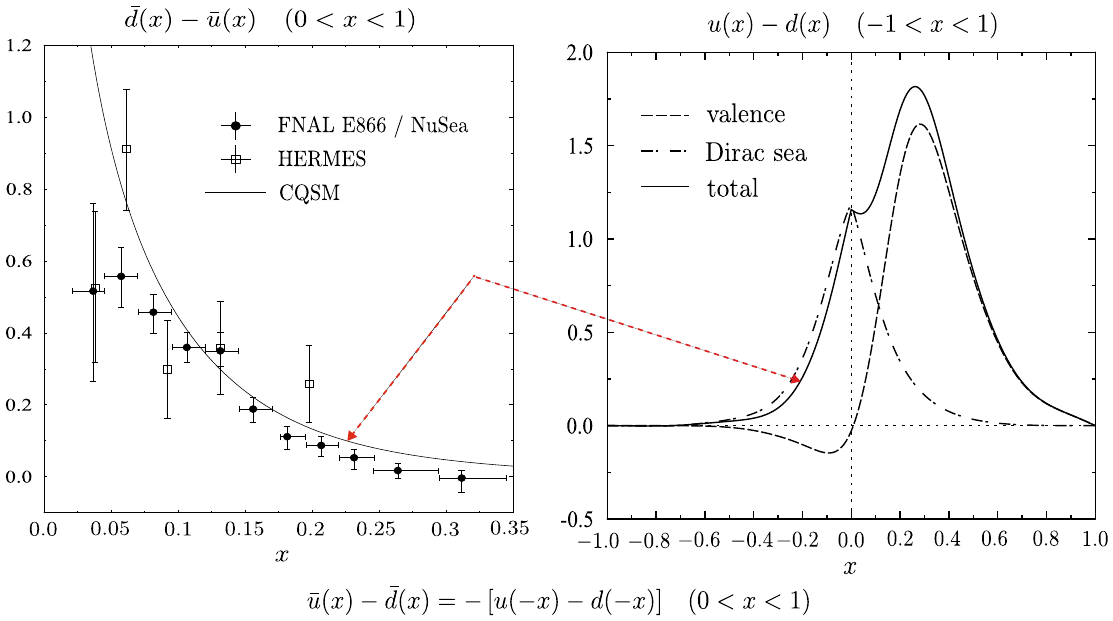}
\caption{Flavor asymmetry of the unpolarized sea-quark (anti-quark)
distributions in the proton. The~left panel shows the prediction of the CQSM
for $\bar{d} (x) - \bar{u} (x)$ in comparison with the experimental data by
the HERMES collaboration~\cite{HERMES1998} and 
the FNAL E866/NuSea collaboration~\cite{E866_NuSea1998}. 
The  prediction of the CQSM here is obtained 
after evolving it to the energy scale of the corresponding experimental data.
For the sake of comparison, the~CQSM prediction at the model energy
scale is re-posted on the right panel. \label{fig07}}
\end{figure}

Now we are in a position to compare the basic predictions of the CQSM for the
twist-2 PDFs evolved to a high energy scale with the corresponding experimental data.
Shown in Figure~\ref{fig07} are the experimental data by the HERMES group
~\cite{HERMES1998}
and the FNAL E866/NuSea group~\cite{E866_NuSea1998}, which clearly show the flavor 
asymmetry of anti-quark distribution in the proton. Undoubtedly, the~$\bar{d}$-quark
distribution dominates over the $\bar{u}$-quark distribution in the proton.
It is known that this flavor asymmetry of the sea-quark distributions can be
explained as a combined effect of the meson clouds and the asymmetry of the
numbers of the $u$-quark and $d$-quark inside the proton. 
(See review~\cite{Kumano1998} for a more detailed explanation.)
In fact, the~following virtual dissociation processes are expected to occur in the proton:
\begin{eqnarray}
 &\,& u \ \rightarrow \ d \ + \ \pi^+, \ \ \ u \ \rightarrow \ u + \pi^0, \\
 &\,& d \ \rightarrow \ u \ + \ \pi^-, \ \ \ d \ \rightarrow \ d + \pi^0.
\end{eqnarray}
Taking account of the quark contents of the pions as
\begin{eqnarray}
 \pi^+  \ &\sim& \ u \,\bar{d}, \\
 \pi^0  \ &\sim& \ \frac{1}{\sqrt{2}} \,(u \,\bar{u} \ - \ d \,\bar{d}), \\
 \pi^-  \ &\sim& \ d \,\bar{u}, 
\end{eqnarray}
one first realizes that the emission of the neutral pion $\pi^0$ generates the same numbers 
of $u \,\bar{u}$ and $d \,\bar{d}$ pairs so that it does not contribute to the asymmetry of
the numbers of $\bar{u}$ and $\bar{d}$ quarks. On~the other hand, the~processes
$u \rightarrow d + \pi^+$ and $d \rightarrow u + \pi^-$ can generate
the difference between the numbers of $u \,\bar{u}$ and $d \,\bar{d}$ pairs.
Since the numbers of the parent $u$-quark and $d$-quark in the proton as the seeds of
this virtual dissociation processes are two and one, this naturally explains
the dominance of the $\bar{d}$-quark distribution over the $\bar{u}$-quark
distribution, at least qualitatively. 
As one sees on the left panel of Figure~\ref{fig07}, the~CQSM reproduces fairly well the
observed difference of the $\bar{d}$-quark and $\bar{u}$-quark distributions without
introducing any adjustable parameters.
This is not surprising, because~the pion cloud effects or the pionic quark--antiquark
excitation modes are automatically included into the model. Undoubtedly, 
such a mechanism is already incorporated in the model
prediction for the distribution $u (x) - d (x)$ given at the low energy model scale.
In order to confirm it, the~CQSM prediction at the model energy
scale is shown again on the right panel of Figure~\ref{fig07}.
As can be demonstrated 
 from this figure, the~contribution of the
negative-energy Dirac-sea quarks have a positive peak with sizable magnitude.
In particular, the~positivity of $u (x) - d (x)$ in the negative $x$ region means
that $\bar{u} (x) - \bar{d} (x)$ for $x > 0$ is negative, which in turn reconfirms that
$\bar{d} (x)$ dominates over $\bar{u} (x)$ in conformity with the observation.   
We can say that the seed of the flavor asymmetry of the anti-quark distribution is 
already contained in the prediction of the CQSM given at the low energy~scale.

\begin{figure}[htbp]
\includegraphics[width=14.5cm]{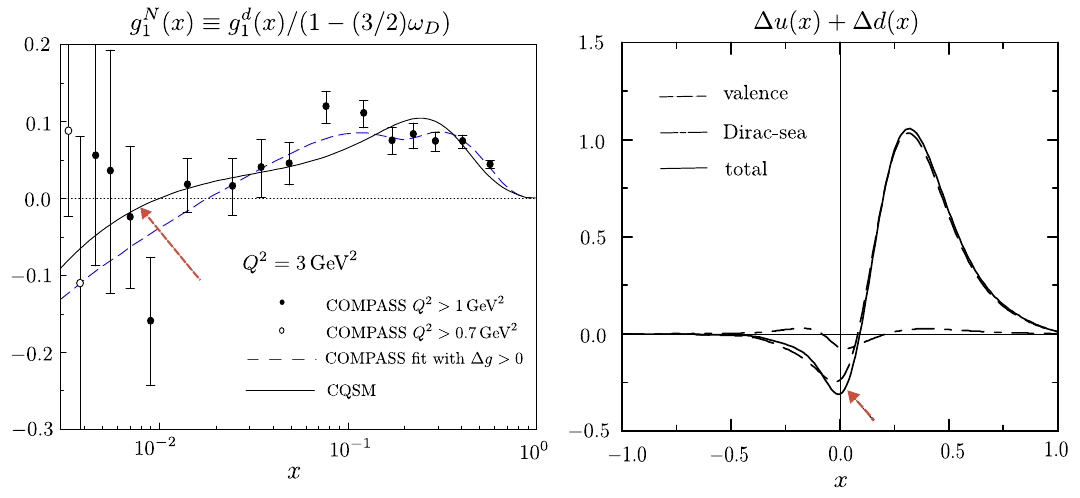}
\caption{The left panel shows the longitudinally polaized distribution function of 
the deuteron $g^N_1 (x) \equiv g^d (x) / ( 1 - (3/2) \,\omega_D)$ 
extracted by the COMPASS group~\cite{COMPASS2005, COMPASS2007} in 
comparison with the prediction of the CQSM for the isoscalar longitudinally polarized 
distribution function $\Delta u(x) + \Delta d (x)$ of the nucleon evolved to the 
corresponding energy scale. Here, $\omega_D \simeq 0.05$ represents the $D$-state 
probability of the deuteron wave function. For~reference, the~CQSM prediction for 
$\Delta u (x) + \Delta d (x)$ at the low energy model scale is re-posted on the right panel.
\label{fig08}}
\end{figure}

\begin{figure}[htbp]
\includegraphics[width=8.5cm]{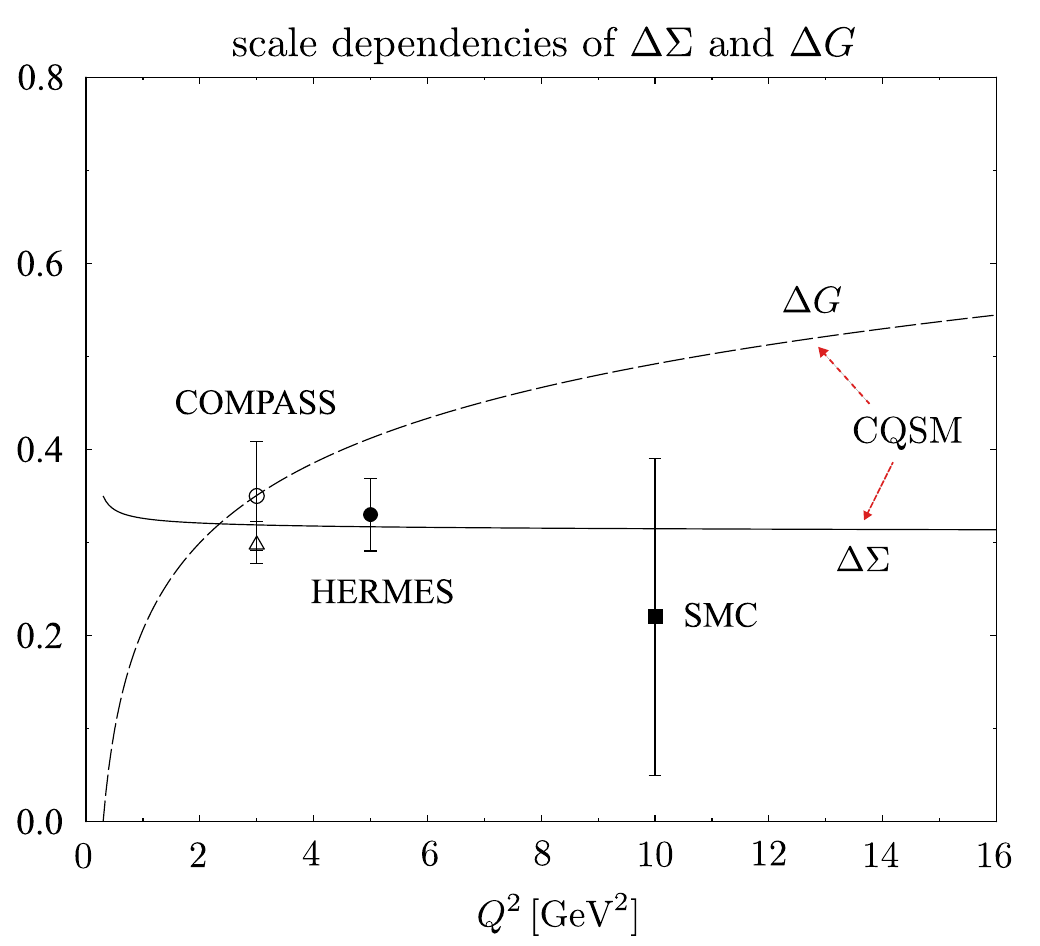}
\caption{Energy  scale dependencies of the quark spin fraction $\Delta \Sigma$
and the gluon spin fraction $\Delta G$ predicted by the CQSM~\cite{WN2006, WN2008},
in comparison with the old fit by the SMC group~\cite{SMC1998} and the newer fits by
the COMPASS group~\cite{COMPASS2005, COMPASS2007} and 
HERMES group~\cite{HERMES2007}. \label{fig09}}
\end{figure}

\vspace{2mm}
Next, we turn our attention to the isoscalar combination of the longitudinally 
polarized distribution, i.e.,~$\Delta u (x) + \Delta d (x)$.
Shown on the left panel of Figure~\ref{fig08} are the empirical data for the
isoscalar longitudinally polarized quark distribution $g^N_1 (x)$ of the nucleon
extracted by the COMPASS group~\cite{COMPASS2005, COMPASS2007}. 
This distribution function is extracted
under the assumption that the deuteron is a weakly bound object of the proton
and the neutron on account of the $D$-state probability $\omega_D \simeq 0.05$ 
in the deuteron wave function. Under~this approximation, the~function $g^N_1 (x)$
can be identified as the isoscalar unpolarized distribution function
$\Delta u (x) + \Delta d (x)$ of the nucleon. 
The corresponding prediction of the CQSM is
shown by the solid curve. One of the  interesting observations 
is that $g^N_1 (x)$ or $\Delta u (x) + \Delta d (x)$ appears to become negative
as $x$ approaches zero, and~this behavior is remarkably consistent with the theoretical
prediction of the CQSM. Interestingly, this behavior is already anticipated
from the prediction of the CQSM at the low energy model scale shown on
the right panel of Figure~\ref{fig09}. As~one sees from this figure, the~contribution
of the Dirac-sea quarks to this distribution function is relatively small.
Rather, the~contribution of the $N_c$ valence quarks determines the   
general tendency of this distribution such that $\Delta u(x) + \Delta d(x) < 0$
in the small $x$ region. To~avoid misunderstanding, we point out that
the valence quark orbital also receives strong deformation under the influence
of the hedgehog mean field. We conjecture that the above-mentioned
peculiar behavior of the valence quark contribution is related to this
strong deformation of the valence quark~orbit.

\vspace{2mm}
The argument above confirms that the prediction of the CQSM for the isoscalar
combination of the longitudinally polarized quark distribution in the nucleon
is consistent with the empirical information extracted by the 
COMPASS group~\cite{COMPASS2005, COMPASS2007} at least qualitatively.
Incidentally, the~first moment of $\Delta u(x) + \Delta d (x)$ is
identified with the net quark spin fraction $\Delta \Sigma$ in the nucleon (it
can be identified with the isoscalar axial-vector coupling constant $g^{(0)}_A$ of 
the nucleon in the gauge-invariant regularization scheme) as
\begin{equation}
 \Delta \Sigma (Q^2) \ = \ \int_{- \,1}^1 \,\left[ \Delta u (x, Q^2) \ + \ 
 \Delta d (x, Q^2) \right] \,d x.    
\end{equation}
Here, we explicitly show the $Q^2$ dependencies of the quantities  
$\Delta \Sigma$ as well as $\Delta u (x)$ and $\Delta d (x)$, which are all
scale-dependent quantities. As~already mentioned, the~CQSM prediction for
$\Delta \Sigma$ at the model energy scale gives
$\Delta \Sigma (Q^2 = 600 \,\mbox{\rm MeV}^2) \simeq 0.35$.
We try to estimate the $Q^2$ dependence of $\Delta \Sigma$ by solving
the coupled evolution equation for $\Delta \Sigma$ and $\Delta G$ at
the next-to-leading order.
For simplicity, the~gluon spin contents $\Delta G$ at the low-energy model
scale is assumed to be negligibly small and it is set to zero. 
Shown in Figure~\ref{fig09} are the scale
dependencies of $\Delta \Sigma$ and $\Delta G$.
As one can see, the~$\Delta \Sigma$ generally has weak scale dependence
at very low energy scales. Above~$Q^2 \geq 1 \,\mbox{GeV}^2$,
it is nearly scale-independent.  Interestingly, the~prediction of the CQSM
for $\Delta \Sigma$ \cite{WN2006, WN2008} looks consistent with all the 
available empirical data, i.e.,~the old SMC data, and~the newer data by the 
HERMES group~\cite{HERMES2007} and COMPASS group~\cite{COMPASS2005, COMPASS2007}.
The evolution equation predicts that, different from $\Delta \Sigma$,  the~gluon 
spin fraction $\Delta G$ has a fairly strong scale dependence and it grows
logarithmically as a function of $Q^2$. 
 
\vspace{-3pt}
  
\vspace{1mm}
Also interesting is the prediction of the CQSM for
the isovector longitudinally polarized quark distribution $\Delta u (x) - \Delta d (x)$
shown in panel $(d)$ of Figure~\ref{fig04}.
Remembering the fact that the distribution $\Delta u (x) - \Delta d (x)$ in the
negative $x$ region can be identified with the anti-quark distribution
$\Delta \bar{u} (x) - \Delta \bar{d} (x)$ with $x > 0$, we see that the
CQSM predicts $\Delta \bar{u} (x) - \Delta \bar{d} (x) > 0$ for the physical value of
$x$ with $0 < x < 1$. Table~\ref{table01} shows the integrals of the distribution
functions $\Delta u (x) + \Delta d (x)$, $\Delta u (x) - \Delta d (x)$, $\Delta u (x)$,
and $\Delta d (x)$ predicted by the CQSM within the range of $x$ designated
in the upper-most row of the table. From~this table, the~first
moments or the integrals of $\Delta \bar{u} (x)$ and $\Delta \bar{d} (x)$ within the
range $0 < x < 1$ can be estimated as
\begin{equation} 
 \Delta \bar{u} \ \simeq 0.092, \ \ \ \Delta \bar{d} \ \simeq \ - 0.139,
\end{equation}
which means that
\begin{equation}
 \Delta \bar{d} \ - \ \Delta \bar{u} \ > \ 0 \ \ \ \mbox{\rm with} \ \ \ 
 \vert \Delta \bar{u} \vert \ \ll \ \vert \Delta \bar{d} \vert .
\end{equation}
At any rate, the~CQSM predicts that the flavor symmetry of the sea-quark distributions
is broken not only for the unpolarized distribution $\bar{u} (x) - \bar{d} (x)$ but also
for the longitudinally polarized distribution $\Delta \bar{u} (x) - \Delta \bar{d} (x)$
in the~proton.


\begin{table}[htbp]
\begin{center}
\caption{The integrals of the longitudinally polarized distribution functions specified 
in the left-most column within the range of $x$ designated in the upper-most row.} 
\label{table01}
\vspace{3mm}
\begin{tabular}{cccc}
\hline
  \textbf{CQSM} \ \ \ & \ \ \ \boldmath{$-1 < x < 0$} \ \ \ & \ \ \ 
  \boldmath{$0 < x < 1$} \ \ \ & \ \ \ \boldmath{$-1< x < 1$}  \ \ \ \\
\hline 
$\Delta u + \Delta d$ & $-$0.047 & 0.399 & 0.352 \\
$\Delta u - \Delta d$ & 0.231 & 1.092 & 1.323 \\
\hline
$\Delta u$ & 0.092 & 0.745 & 0.838 \\
$\Delta d$ & $-$0.139 & $-$0.346 & $-$0.485 \\
\hline
\end{tabular}
\end{center}
\end{table}

\begin{figure}[htbp]
\includegraphics[width=15.0cm]{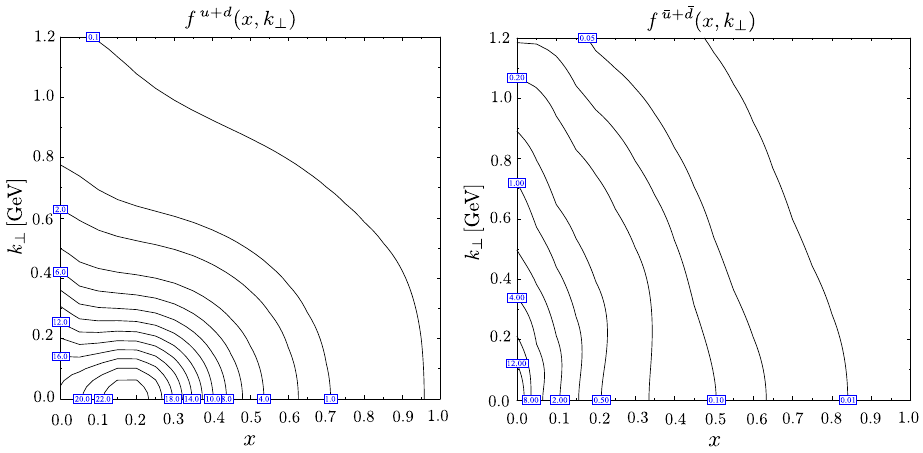}
\caption{Contour plot of the isoscalar unpolarized transverse momentum-dependent (TMD)
quark distribution functions predicted by the CQSM.
The left panel represents the TMD quark distribution 
$f^{u+d} (x, k_\perp) \equiv u (x, k_\perp) + d (x, k_\perp)$,
while the right panel does the TMD anti-quark distribution 
$f^{\bar{u}+\bar{d}} (x, k_\perp) \equiv \bar{u} (x, k_\perp) + \bar{d} (x, k_\perp)$.\label{fig10}}
\end{figure}

Before ending this section, let us briefly touch upon the CQSM prediction for the
transverse momentum-dependent (TMD) distribution function of the nucleon~\cite{Waka2009}.
Shown in Figure~\ref{fig10} is the contour plot for the isoscalar unpolarized TMD quark 
distribution function. The~left panel stands for the quark distribution
$f^{u + d} (x, \bm{k}_\perp) \equiv u (x, k_\perp) + d (x, k_\perp)$, while the right panel does 
the corresponding anti-quark distribution 
$f^{\bar{u} + \bar{d}} (x, \bm{k}_\perp) \equiv \bar{u} (x, k_\perp) + \bar{d} (x, k_\perp)$.
Although it is not very easy to see only from these figures, the~above predictions
of the CQSM indicate that the frequently assumed factorized form with the
Gaussian distribution in $k_\perp$ given as
\begin{equation}
 f^{u+d} (x, \bm{k}_\perp)  \ = \ f^{u+d} (x) \,\frac{1}{\pi \,\langle k^2_\perp \rangle}
 \,e^{\,- \,k^2_\perp / \langle k^2_\perp \rangle} \ \ \ \mbox{\rm with} \ \ 
 \langle k^2_\perp \rangle \simeq \ 0.25 \,\mbox{\rm GeV}^2 ,
\end{equation}
is not necessarily~justified.

\begin{figure}[htpb]
\includegraphics[width=8.0cm]{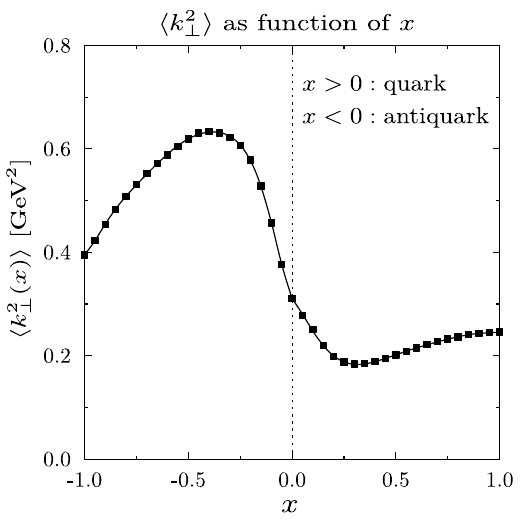}
\caption{The CQSM prediction for the average transverse momentum square of
quarks ($x > 0$) and of antiquarks ($x < 0$) as a function of the longitudinal
momentum fraction $x$.\label{fig11}}
\end{figure}

\vspace{2mm}
To confirm the above statement more clearly, we evaluate from the theoretical
distribution $f^{u + d} (x, \bm{k}_\perp)$ the average transverse momentum of quarks 
and anti-quarks as functions of $x$, which is defined by
\begin{equation}
 \langle k^2_\perp (x) \rangle \ \equiv \ 
 \frac{\int \,d^2 \bm{k}_\perp \,k^2_\perp \,f^{u+d} (x,\bm{k}_\perp)}{\int \,d^2 \bm{k}_\perp \,
 f^{u+d} (x, \bm{k}_\perp)} .
\end{equation}
The resultant $\langle k^2_\perp (x) \rangle$ is shown by filled squares in Figure~\ref{fig11}.
The solid curve here is a smooth fit to the above numerical results by an 8th-order 
polynomial.
One clearly sees that the average transverse momentum square is strongly dependent
on the longitudinal momentum fraction $x$ of quarks and anti-quarks, which
confirms that the frequently used assumption of factorization in the variables
$x$ and $k_\perp$ is significantly broken.
Also worthy of special mention here is very unique prediction of the CQSM,
which indicates that the magnitude $\langle k^2_\perp (x) \rangle$ is much larger
in the negative $x$ region corresponding to the anti-quarks.
It can be demonstrated 
 more clearly by comparing the two quantities defined below:  
\vspace{+6pt}
\begin{eqnarray}
 \langle k^2_\perp \rangle^Q \ &\equiv& \ 
 \frac{\int_0^1 \,d x \,\langle k^2_\perp (x) \rangle \,f^{u+d} (x)}
 {\int_0^1 \,d x \,f^{u+d} (x)}, \\
 \langle k^2_\perp \rangle^{\bar{Q}} \ &\equiv& \ 
 \frac{\int_ {- \,1}^0 \,d x \,\langle k^2_\perp (x) \rangle \,f^{u+d} (x)}
 {\int_{- \,1}^0 \,d x \,f^{u+d} (x)} \ = \ 
 \frac{\int_0^1 \,d x \,\langle k^2_\perp (x) \rangle \,f^{\bar{u} + \bar{d}} (x)}
 {\int_0^1 \,d x \,f^{\bar{u} + \bar{d}} (x)}, 
\end{eqnarray}
which represents the average transverse momentum square for quarks and anti-quarks,
respectively. Numerically, we find that
\begin{eqnarray}
 \langle k^2_\perp \rangle^Q \ &=& \ 0.224 \,\mbox{\rm GeV}^2, \\
 \langle k^2_\perp \rangle^{\bar{Q}} \ &=& \ 0.445 \,\mbox{\rm GeV}^2,
\end{eqnarray}
which shows that the average transverse momentum of anti-quarks is
sizably larger than that of quarks.
We emphasize that what makes this nontrivial prediction possible is the prominent 
nature of the CQSM in which one is able to take account of nonperturbative deformation 
of the negative-energy Dirac-sea orbitals in the hedgehog mean~field.

\vspace{-3pt}

\section{Flavor SU(3) CQSM and Strange-Sea Asymmetry in the~Nucleon}
\label{sec:SU(3)_CQSM}

It is believed that the quantum nucleon state has, in~general, nonzero strange
quark components. To~take such possibilities into account,
we need a flavor SU(3) extension of the CQSM. 
The flavor SU(3) version of the CQSM is characterized by the following effective
Lagrangian~\cite{BDGPPP1993, ARW1996}
\begin{equation}
 {\cal L} \ = \ \bar{\psi} (x) \,\left( \,i \,\slashed{\partial} \ -\ M \,U^{\gamma_5} (x)
 \ - \ \Delta m_s \,P_s \right) \,\psi (x), \label{Eq:SU(3)_CQSM}
\end{equation}
where
\begin{equation}
 U^{\gamma_5} (x) \ = \ e^{\,i \gamma_5 \,\pi (x) / f_\pi} \ \ \ \mbox{\rm with}
 \ \ \ \pi (x) \ = \ \pi_a (x) \,\lambda_a \ \ \ ( a = 1, \cdots, 8 )
\end{equation}
Here, $\pi (x)$ stands for the octet meson fields.
The last term in the parenthesis on the r.h.s of Equation~(\ref{Eq:SU(3)_CQSM}) represents 
the SU(3) breaking term, which is given as
\begin{equation}
 \Delta m_S \, P_s \ = \
 \left( \begin{array}{ccc}
 0 & \ \ 0 \ \ & 0 \\
 0 & \ \ 0 \ \ & 0 \\
 0 & \ \ 0 \ \ & \Delta m_s \\
 \end{array}
 \right), 
\end{equation}
with $\Delta m_s$ being the mass difference between the strange quark and
the non-strange quarks. In~the following, we set $\Delta m_s = (80$$\,\sim \,$$100) \,\mbox{\rm MeV}$.

\vspace{2mm}
The basic dynamical assumptions of the flavor SU(3) CQSM are as follows:
\begin{itemize} 
 \item[(1)] The lowest energy classical solution is obtained by embedding the SU(2)
 hedgehog solution to the SU(3) matrix as follows:
\vspace{+6pt}
\begin{equation}
  U^{\gamma_5}_0 (\bm{r}) \ = \ \left( \begin{array}{cc}
  e^{\,i \,\gamma_5 \,\bm{\tau} \cdot \hat{\bm{r}} \,F (r)} & 0 \\
  0 & 1 \\
  \end{array} \right) .
 \end{equation}
 \item[(2)] Quantization of soliton rotational motion in the SU(3) collective coordinate space
 \item[(3)] Perturbative treatment of the SU(3) breaking term given by
\begin{equation}
 \Delta \tilde{H} \ = \ \Delta m_s \,\,\gamma^0 \,A^\dagger (t) \,
 \left( \frac{1}{3} \ - \ \frac{1}{\sqrt{3}} \,\lambda_8 \right) \,A (t) .
 \end{equation}
\end{itemize}

The comparison with the high-energy data is carried out similarly with the case of
SU(2) CQSM. We use the predictions of the CQSM for~\cite{Waka2003A, Waka2003B}
\begin{eqnarray*}
 &\,& u (x), \,d (x), \,s (x) \ \ \ \mbox{\rm and} \ \ \ \Delta u(x), \,\Delta d (x), \,\Delta s (x), \\
 &\,& \bar{u} (x), \,\bar{d} (x), \,\bar{s} (x), \ \ \mbox{\rm and} \ \ \ 
 \Delta \bar{u} (x), \,\Delta \bar{d} (x), \,\Delta \bar{s} (x), 
\end{eqnarray*}
as initial-scale quark and anti-quark distributions given at the model energy scale 
\linebreak  $Q^2_{ini} \simeq ( 600 \,\mbox{\rm MeV})^2$.
The gluon distribution functions at this low energy scale is simply set to be zero as before,
\[
 g (x) \ = \ 0 \ \ \mbox{\rm and} \ \ \Delta g (x) \ = \ 0.
\]
(There is some phenomenological indication, which implies that 
the above assumption for the longitudinally polarized gluon distribution at the above low
energy scale  $\Delta g (x)$ is not so bad but the unpolarized gluon distribution 
$g (x)$ at the same energy scale is not necessarily negligible.)

\vspace{2mm}
The greatest advantage of the SU(3) CQSM is that it can give nontrivial prediction for
the asymmetry of the strange and anti-strange quark distributions in the proton.
Since the net strange quark number in the proton is zero, there is a rigorous
sum rule between the unpolarized strange quark distribution function $s (x)$ and
the corresponding anti-strange quark distribution $\bar{s} (x)$: 
\begin{equation}
 \int_0^1 \,\left[ s (x) \ -\ \bar{s} (x) \right] \,d x \ = \ 0.
\end{equation}
However, this does not necessarily dictate that the distributions $s (x)$ and $\bar{s} (x)$ 
are identical. 
In fact, in~the physical picture of the meson--baryon fluctuation model due to
~\cite{Sullivan1972, ST1987}, 
the following virtual dissociation process of the proton is expected to occur,
\begin{equation}
 p \ \rightarrow \ \Lambda \ + \ K^+. \label{Eq:Kaon_cloud}
\end{equation}
We recall here the quark contents of $\Lambda$ and $K^+$, which are given as
$\Lambda \sim u d s$ and $K^+ \sim u \bar{s}$.
Note that on the r.h.s. of (\ref{Eq:Kaon_cloud}), the~$s$-quark is contained
in $\Lambda$ (baryon), while the $\bar{s}$-quark is contained in $K^+$ (meson).
This already indicates that the distribution functions for the $s$-quark and
the $\bar{s}$-quark need not be the same. That is, the~$s$-quark is expected 
to have valence-like harder component as compared with the $\bar{s}$-quark
contained in the soft meson. (To have a harder component means that it prevails in a 
larger $x$ region.) This in turn indicates that the enhancement of the difference
distribution $s (x) - \bar{s} (x)$ at larger $x$.

\begin{figure}[htpb]
\includegraphics[width=11.0cm]{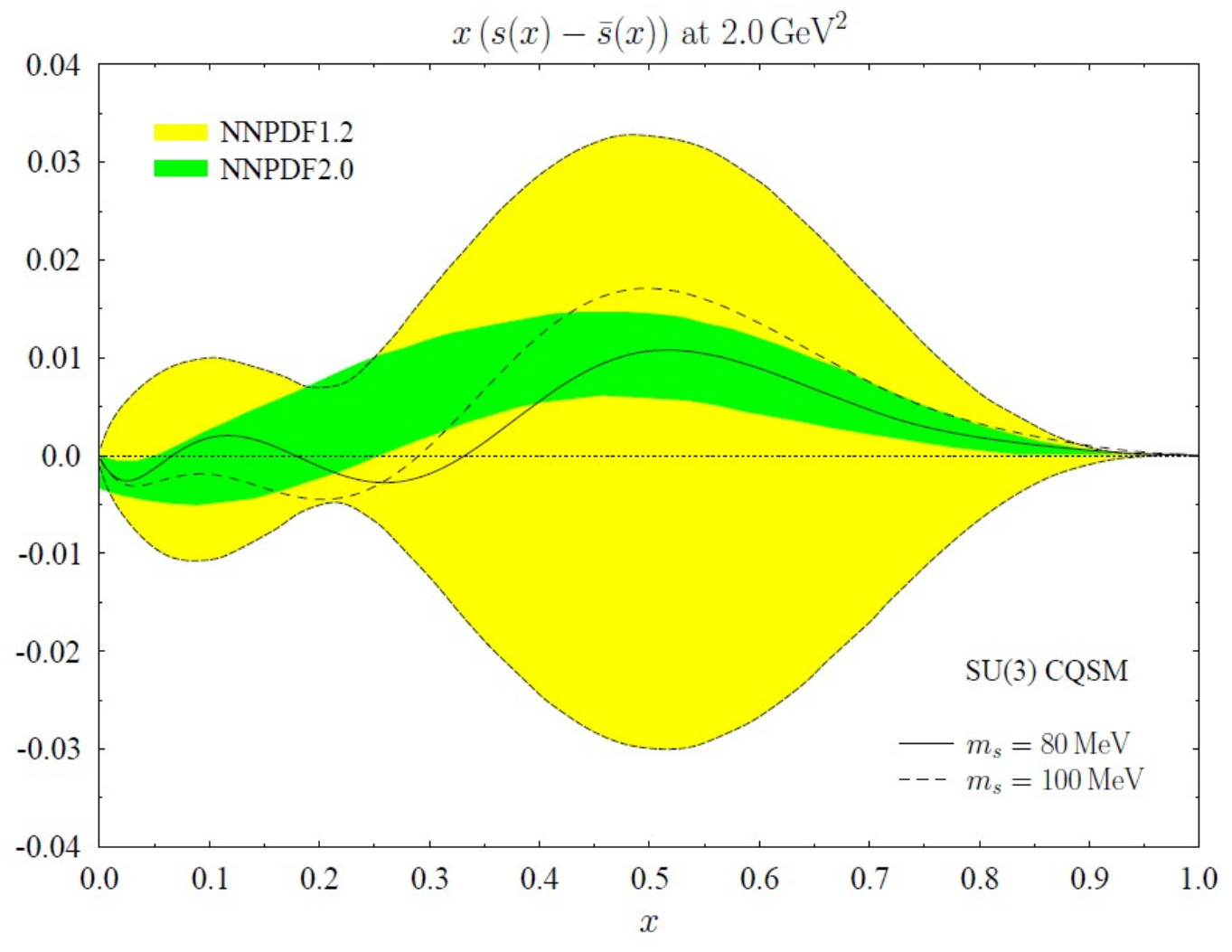}
\caption{The prediction of the SU(3) CQSM for the unpolarized distribution
$x \,( s(x) - \bar{s} (x))$ inside the nucleon in comparison with the global fit 
by the NNPDF collaboration~\cite{NNPDF2010, NNPDF2012}.}
\label{fig12}
\end{figure}

We show in Figure~\ref{fig12} the prediction of the SU(3) CQSM for the distribution
$x \,( s(x) - \bar{s} (x))$ in comparison with the global fit by the NNPDF 
collaboration~\cite{NNPDF2010, NNPDF2012}.
The model prediction, i.e.,~the enhancement of $s (x)$ over $\bar{s} (x)$ in the
larger $x$ region, appears to be consistent with the newer fit NMPDF2.0
(and NNPDF2.1), at least~qualitatively.

\begin{figure}[htpb]
\includegraphics[width=8.5cm]{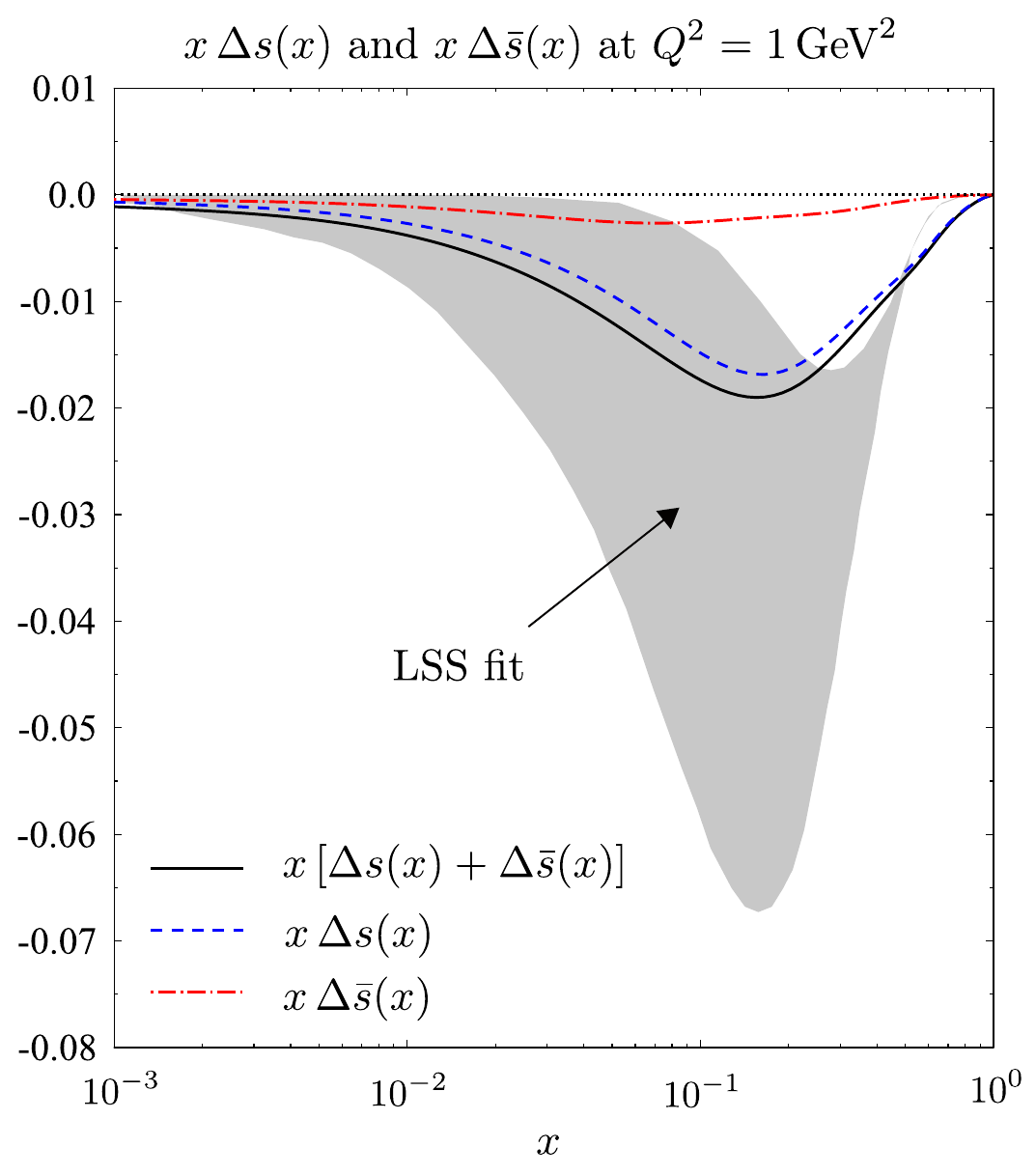}
\caption{The prediction of the SU(3) CQSM for the differences between the 
longitudinally polarized strange and anti-strange distributions inside the nucleon
in comparison with the LSS fit for the average distribution
$x \,[\Delta s (x) + \Delta \bar{s} (x)]$ \cite{LSS2003, LSS2006}.}
\label{fig13}
\end{figure}

A question is whether one can expect asymmetry also for the longitudinally
polarized strange and anti-strange quark distributions.
Many years ago, Brodsky and Ma~\cite{BM1996} predicted
the asymmetry of the distribution functions $\Delta s (x)$ and $\Delta \bar{s} (x)$
based on the kaon cloud model of the nucleon. They argued that the strange
sea in the proton is thought to be generated through the virtual dissociation
process of the proton into $\Lambda$ and $K^+$, $p \rightarrow \Lambda + K^+$. 
Note that here is an apparent
asymmetry of the $s$-quark and $\bar{s}$-quark in this process. 
The $s$-quark is contained in the spin one-half baryon, i.e.,~$\Lambda$, 
while $\bar{s}$-quark is contained in the spin zero meson, i.e.,~$K^+$. 
Since the polarization of $\bar{s}$-quark in the spin zero $K^+$ is zero on the 
average, they conjectured that the polarization of $\bar{s}$-quark would be smaller 
than that of $s$-quark inside the parent proton.
Shown in Figure~\ref{fig13} is the prediction of the SU(3) CQSM for the distributions
$x \,s (x)$, $x \,\bar{s} (x)$, and~$x \,[ \Delta s(x) + \Delta \bar{s} (x)]$ evolved to
the energy scale of $Q^2 = 1 \,\mbox{\rm GeV}^2$. The~corresponding phenomenological
fit for $x \,[ \Delta s(x) + \bar{s} (x)]$ by the LSS group is also shown
for reference~\cite{LSS2003, LSS2006}. 
One sees that the prediction of the SU(3) CQSM confirms the
conjecture by Brodsky and Ma. The~magnitude of the distribution 
$\Delta \bar{s} (x)$ is seen to be much smaller than that of $\Delta s (x)$.
This confirms that effects of kaon cloud is taken into account automatically 
and effectively in the framework of SU(3) CQSM.
Owing to the difference of the dynamical assumptions, the~SU(3) CQSM gives
somewhat different predictions from the SU(2) CQSM, even for some of the
non-strange $u$-quark and $d$-quark~distributions.

\begin{figure}[htpb]
\includegraphics[width=9.0cm]{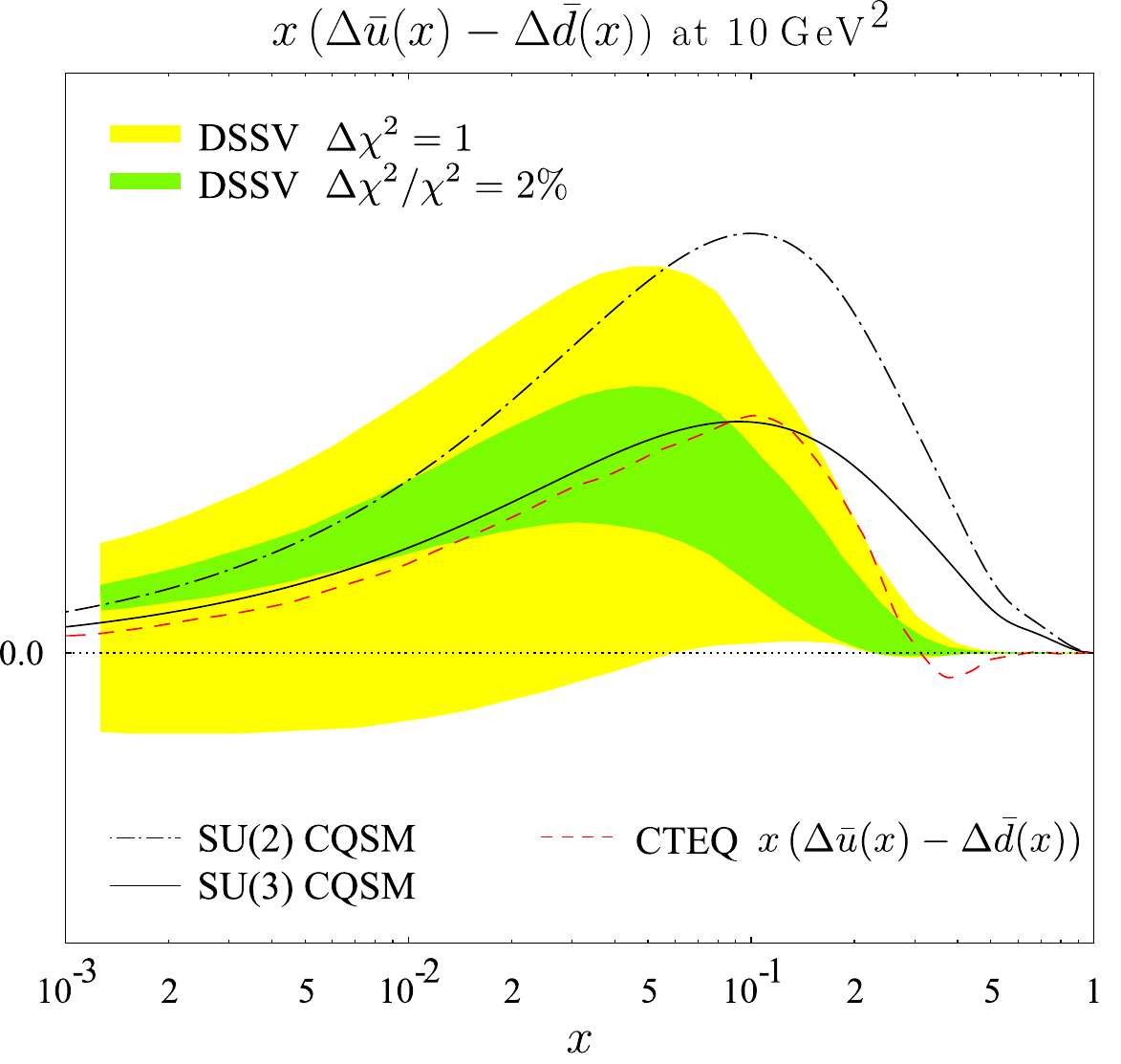}
\caption{The comparison of the predictions of the SU(3) CQSM and the SU(2) CQSM
for the differences between the longitudinally polarized $\bar{u}$-quark and
$\bar{d}$-quark distributions inside the proton in comparison with the DSSV
fit~\cite{DSSV2009} given at the energy scale of $Q^2 = 10 \,\mbox{\rm GeV}^2$.}
\label{fig14}
\end{figure}

\vspace{2mm}
We show in Figure~\ref{fig14} the comparison of the predictions of the SU(3) CQSM 
and the SU(2) CQSM for the difference distribution
$x \,(\Delta \bar{u} (x) - \Delta \bar{d} (x))$ in the nucleon.
Also shown in this figure is the DSSV fit~\cite{DSSV2009} for the same distribution given
at the energy scale of $Q^2 = 10 \,\mbox{\rm GeV}^2$.
One sees that both of the SU(2) CQSM and the SU(3) CQSM predict sizable flavor 
asymmetry of the longitudinally polarized $\bar{u}$-quark and $\bar{d}$-quark
distributions. However, the~magnitude of the flavor asymmetry in the SU(3) CQSM
is seen to be sizably suppressed as compared with the prediction of the SU(2) CQSM.
The reason may be interpreted as follows. 
We recall that one of the basic theoretical postulates in the SU(3) CQSM
is the collecive rotation of the hedgehog object in the flavor SU(3) space.
This collective rotation is expected to drive away some of
the pion clouds into the kaon clouds. This is equivalent to saying that
some of the $u$- and $d$-quark components are driven away to
the strange quark sector. It is thought to explain the possible reason of   
the reduction of the $x \,(\Delta \bar{u} (x) - \Delta \bar{d} (x))$ in the SU(3) 
CQSM as compared with that in the SU(2) CQSM, as illustrated in Figure~\ref{fig14}.   
At any rate, it is interesting to
see that the prediction of the SU(3) CQSM for the size of the difference distribution 
$x \,( \Delta \bar{u} (x) - \Delta \bar{d} (x))$ looks qualitatively consistent with
that of the DSSV global fit~\cite{DSSV2009}.  

\vspace{-6pt}
\vspace{-12pt}

\begin{figure}[htpb]
\includegraphics[width=14.0cm]{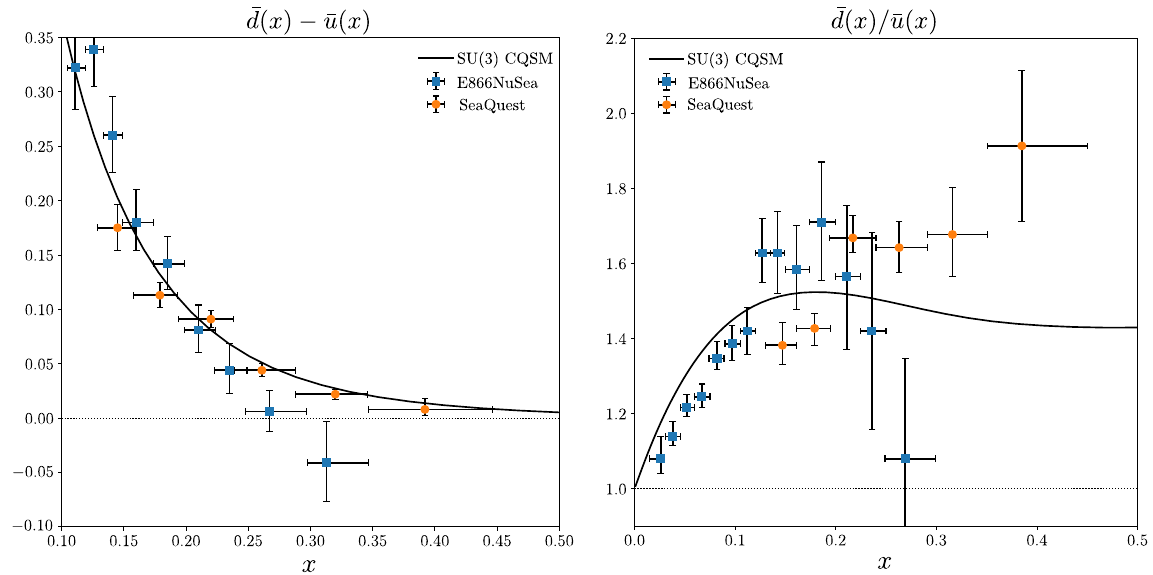}
\caption{The prediction of the SU(3) CQSM for the difference and the ratio
of the unpolarized $\bar{d}$-quark and $\bar{u}$-quark distributions inside the proton
in comparison with the latest SeaQuest fit~\cite{SeaQuest2025} and the 
past E866/NuSea fit~\cite{E866_NuSea1998}.}
\label{fig15}
\end{figure}

Finally, we show in Figure~\ref{fig15} the predictions of the SU(3) CQSM for the difference
and the ratio of the unpolarized $\bar{d}$-quark and $\bar{u}$-quark distributions in 
the proton in comparison with the latest SeaQuest fit~\cite{SeaQuest2025} together with 
the past E866/NuSea fit~\cite{E866_NuSea1998}.
The main difference between the newest SeaQuest data and the old E866/NuSea data
for the difference distribution is its behavior in larger $x$ region.
The old E866/NuSea fit shows that the distribution $\bar{d} (x) - \bar{u} (x)$
changes its sign around $x \simeq (0.25 \sim 0.3)$. 
On the other hand, according to the new 
SeaQuest fit, $\bar{d} (x) - \bar{u} (x)$ remains positive up to the larger $x$ range. 
The latter behavior was the feature expected from most effective models of the nucleon including
the CQSM. In any case, the~prediction of the SU(3) CQSM looks fairly consistent with the 
new SeaQuest~fit.

A far more delicate quantity is the ratio of the $\bar{d}$- and $\bar{u}$-quark
distributions in the larger $x$ region. This is because, in~the ratio
$\bar{d} (x) / \bar{u} (x)$ at larger $x$, the~denominator and the numerator
are both very small quantities. This makes the experimental extraction of 
the ratio  $\bar{d} (x) / \bar{u} (x)$ at larger $x$ very hard.
On the right panel of Figure~\ref{fig15}, new SeaQuest fit~\cite{SeaQuest2025} and the 
old E866/NuSea fit~\cite{E866_NuSea1998} for the ratio 
 $\bar{d} (x) / \bar{u} (x)$ are 
shown together with the corresponding
prediction of the SU(3) CQSM. As~anticipated from the sign change of the
difference distribution $\bar{d} (x) - \bar{u} (x)$ around $x \simeq (0.25 \sim 0.3)$, 
the old E866/NuSea fit indicates that the ratio $\bar{d} (x) / \bar{u} (x)$ becomes
smaller than unity around the same value of $x$. On~the other hand, the~new
SeaQuest fit shows that this ratio remains larger than unity or even becomes
much larger than unity as $x$ increases. It seems to us that this drastic discrepancy
between the new and old extraction of the ratio $\bar{d} (x) / \bar{u} (x)$ indicates 
the hardness of reliable extraction of this quantity. At~the present stage, 
we can just say that the prediction of the SU(3) CQSM lies between the new and 
old experimental~extractions.

\section{On the Gauge-Invariant Decomposition Problem of the Nucleon~Spin}
\label{sec:GI_decomposition}

It has been long known that there exist two different decompositions of the
nucleon spin. The~one is the Ji decomposition~\cite{Ji1997}, while the other is the Jaffe--Manohar
decomposition~\cite{JM1990}. The~Ji decomposition is symbolically expressed as
\begin{equation}
 \bm{J}_{QCD} \ = \ \bm{S}^Q \ + \ \bm{L}^Q \ + \ \bm{J}^G ,
\end{equation}
where
\vspace{+6pt}
\begin{eqnarray}
 \bm{S}^Q \ &=& \ \frac{1}{2} \,\int \,\psi^\dagger \,\bm{\Sigma} \,\psi \,d^3 x, \\
 \bm{L}^Q \ &=& \ \int \,\psi^\dagger \,\bm{x} \times \frac{1}{i}\,\bm{D} \,\psi \,d^3 x, \\
 \bm{J}^G \ &=& \ \int \,\bm{x} \times \left(\bm{E}^a \times \bm{B}^a \right) \,d^3 x .
\end{eqnarray}
Here, $\bm{D} = \nabla - i \,g \,\bm{A}$ represents the standard covariant derivative,
while $\bm{S}^Q$ and $\bm{L}^Q$ with $Q = u + d + \cdots$ respectively represent 
the contributions of the intrinsic spin and the orbital angular momentum (OAM) of quark fields. 
Note that the quark OAM $\bm{L}^Q$ appearing in this decomposition is the manifestly 
gauge-invariant mechanical quark OAM
as emphasized in~\cite{Waka2010, Waka2011}.
According to Ji,  the~total gluon angular momentum $\bm{J}^Q$ of the gluon cannot be 
gauge-invariantly decomposed into the contributions
of the intrinsic spin and orbital angular momentum parts~\cite{Ji1997}.  

\vspace{2mm}
On the other hand, the~Jaffe--Manohar decomposition is given as
\begin{equation}
 \bm{J}_{QCD} \ =\ \bm{S}^Q \ + \ \bm{L}^{\prime Q} \ + \ \bm{S}^G \ + \ \bm{L}^G ,
\end{equation}
where
\begin{eqnarray}
 \bm{S}^Q \ &=& \ \frac{1}{2} \,\int \,\psi^\dagger \,\bm{\Sigma} \,\psi \,d^3 x, \\
 \bm{L}^{\prime Q} \ &=& \ \int \,\psi^\dagger \,\bm{x} \times \frac{1}{i} \,\nabla \,\psi \,d^3 x, \\
 \bm{S}^G \ &=& \ \int \,\bm{E}^a \times \bm{A}^a \,d^3 x, \\
 \bm{L}^G \ &=& \ \int \,E^{a i} \,\bm{x} \times \nabla \,A^{a i} \,d^3 x,
\end{eqnarray}
where $a = 1, \cdots, 8$ represents the color quantum numbers.
The quark OAM $\bm{L}^{\prime Q}$ appearing in this decomposition is 
the so-called canonical quark OAM, i.e.,~$\bm{L}^{\prime Q} = \bm{L}^Q_{can}$.
It had been long believed that, in~this Jaffe--Manohar decomposition of the nucleon spin,
only the intrinsic quark spin part $\bm{S}^Q$ is
gauge-invariant, while other three pieces are all gauge-dependent quantities.
However, after~Chen~et~al.'s paper appeared~\cite{CLSWG2008, CSLWG2009}, 
several authors proposed the concept of gauge-invariant extension of the canonical 
OAM~\cite{Hatta2011, Lorce2013}, and~the belief that the canonical OAM extended
in that way can be regarded as a gauge-invariant quantity 
became popular. (See the reviews~\cite{LL2014, Waka2014} for the lively debate at that time.)
The theoretical basis of this idea is the decomposition of the vector 
potential $\bm{A}$ into its physical and pure-gauge component, 
$\bm{A} = \bm{A}_{phys} + \bm{A}_{pure}$ \cite{CLSWG2008, CSLWG2009}.
An apparent problem of such an idea is that the way of decomposition of the vector
potential into the two components is not unique, so there are plural possibilities of 
extension. Upon~noticing such frustrating circumstances, Ji, Xu, and~Zhao advocated the
viewpoint~\cite{JXZ2012} that the Chen decomposition is a gauge-invariant extension of the Jaffe--Manohar 
decomposition based on the Coulomb gauge~\cite{CLSWG2008, CSLWG2009}, 
while the Bashinsky--Jaffe decomposition is a gauge-invariant extension of the 
Jaffe--Manohar decomposition based on the light-cone gauge 
~\cite{BJ1998}.
According to them, since the ways of gauge-invariant extension are not  unique, 
there is no need for the two decompositions to yield the same physical predictions. 
This makes Ji revive his longstanding claim that the gluon spin
$\bm{S}^G$ (or $\Delta G$) has a meaning only in the light-cone gauge, and~it is
not a gauge-invariant quantity in a true or traditional sense. 
Probably, this is a statement that captures the core of the problem.
If that claim is true, however, the~terminology gauge-invariant extension does not 
reflect the truth, because~the so-extended canonical OAM and/or gluon spin are actually 
gauge-variant quantities. Undoubtedly, it is better to use the term gauge-covariant extension 
instead of the term gauge-invariant extension. The~critical difference between these two 
terminologies will be clarified~shortly. 

In fact, to~satisfactorily judge whether some quantity is gauge-invariant or not is a fairly
hard challenge in QCD. The~reason is that, to~show the gauge invariance of  some
quantity, one must evaluate and compare the expectation values of the corresponding 
operator between the system eigen-states in different gauges. 
Unfortunately,  we have no analytical tool in QCD to obtain 
the exact nucleon eigen-states.  
To gain some clear insight into this difficult problem, we proposed to consider an 
intimately connected but much simpler problem in quantum mechanics~\cite{WH2022, WKZ2018}, that is the familiar Landau problem, which handles
the quantum mechanical motion of an electron in a uniform magnetic field. 
The key in this investigation is the quantity 
called the pseudo angular momentum in the Landau electron system~\cite{KM2016, KM2017}.
It is known that, under~the presence of the uniform magnetic field $\bm{B}$ 
directed to the $z$-direction, there exist three types of orbital angular momenta.
In the cylindrical coordinate system $\bm{x} = (r, \,\phi)$ with $r = \sqrt{x^2 + y^2}$
and $\phi = \arctan \left( \frac{y}{x} \right)$, they are expressed as
\begin{eqnarray}
 \hat{L}^{can}_z \ &=& \ - \,i \,\frac{\partial}{\partial \phi} \,, \\
 \hat{L}^{mech}_z (\bm{A}) \ &=& \ - \,i \,\frac{\partial}{\partial \phi} \ + \ e \,r \,A_\phi \,, \\
 \hat{L}^{ps}_z (\bm{A}) \ &=& \ - \,i \,\frac{\partial}{\partial \phi} \ + \ e \,r \,A_\phi
 \ - \ \frac{1}{2} \,e \,B \,r^2 .
\end{eqnarray}
Here, $\hat{L}^{can}_z$ and $\hat{L}^{mech}_z (\bm{A})$ are the familiar canonical and
mechanical angular momentum operators, while $\hat{L}^{ps}_z (\bm{A})$ is called the
pseudo angular momentum operator~\cite{KM2016, KM2017} or the conserved angular 
momentum operator in the terminology of~\cite{WH2022}. 
Suppose that \mbox{$U (\bm{x}) = e^{\,i \,e \,\chi (\bm{x})}$} is a $U(1)$ gauge transformation
matrix which generates the transformation of the vector potential from $\bm{A} (\bm{x})$
to $\bm{A}^\prime (\bm{x})$.
Under this gauge transformation, the~mechanical OAM operator $\hat{L}^{mech}_z (\bm{A})$ 
transforms gauge-covariantly, i.e.,
\begin{equation}
 \hat{L}^{mech}_z (\bm{A}) \ \rightarrow \ \hat{L}^{mech}_z (\bm{A}^\prime) \ = \ 
 U (\bm{x}) \, \hat{L}^{mech}_z (\bm{A}) \,U^\dagger (\bm{x}) .
\end{equation}
Interestingly enough, the~pseudo OAM operator $\hat{L}^{ps}_z$ also transforms
gauge-covariantly as
\begin{equation}
 \hat{L}^{ps}_z (\bm{A}) \ \rightarrow \ \hat{L}^{ps}_z (\bm{A}^\prime) \ = \ 
 U (\bm{x}) \, \hat{L}^{ps}_z (\bm{A}) \,U^\dagger (\bm{x}) .
\end{equation}
It can be easily demonstrated 
 as follows. First, note that the pseudo OAM operator above can be
expressed as $\hat{L}^{ps}_z (\bm{A}) = \hat{L}^{mech}_z (\bm{A}) - \frac{1}{2} \,e \,B \,r^2$.
Here, $\hat{L}^{mech}_z (\bm{A})$ transforms gauge-covariantly, while term
$- \,\frac{1}{2} \,e \,B \,r^2$ is obviously intact under a gauge transformation. 
This means that $\hat{L}^{ps}_z$ also transforms gauge-covariantly.  
Also noteworthy is the following fact. In~the so-called symmetric gauge choice
of the vector potential $\bm{A}^{(S)} (\bm{x}) = \frac{1}{2} \,B \,( - \,y, x)$,
the pseudo OAM operator just reduces to the ordinary canonical OAM operator,
\begin{equation}
 \hat{L}^{ps}_z (\bm{A}) \ \stackrel{\bm{A} \rightarrow \bm{A}^{(S)}}{\longrightarrow} \ 
 \hat{L}^{can}_z.
\end{equation}
This implies that the pseudo OAM operator can be interpreted as a gauge-covariant
extension of the canonical OAM operator based on the symmetric gauge.
From the gauge-covariant transformation property of $\hat{L}^{mech}_z (\bm{A})$ 
and $\hat{L}^{ps}_z (\bm{A})$, one might be tempted to conclude that both correspond 
to gauge-invariant observables. This is not true, however. 
There actually exists a vital difference between the
mechanical and pseudo OAMs. To~see it, we first recall that there are two other typical
choices of gauge in the Landau problem, which are called the 1st and 2nd Landau gauges.
The corresponding gauge potentials in the three gauge choices are characterized as
\vspace{+6pt}
\begin{equation}
 \bm{A}^{(S)} (\bm{x}) \ = \ \frac{1}{2} \,B \,\left( - \,y, x \right), \ \ \ 
 \bm{A}^{(L_1)} (\bm{x}) \ = \ B \left( 0, x \right), \ \ \ 
 \bm{A}^{(L_2)} (\bm{x}) \ = \ B \left(- \,y, 0 \right).
\end{equation}
The eigen-states in the symmetric gauge are usually denoted as $\vert \Psi^{(S)}_{n, m} \rangle$,
where $n$ is the familiar Landau quantum number, while $m$ is the eigen-value
of the canonical OAM operator $\hat{L}^{can}_z$ or the pseudo OAM operator $\hat{L}^{ps}_z (\bm{A})$.
On the other hand, the~eigen-states in the 1st Landau gauge are represented as
$\vert \Psi^{(L_1)}_{n, k_x} \rangle$, where $n$ is the Landau quantum number again, while
$k_x$ is the eigen-value of the canonical momentum operator $\hat{p}^{can}_x$ or
the pseudo momentum operator $\hat{p}^{ps}_x (\bm{A})$. (We do not repeat the analogous
explanation of the 2nd Landau gauge eigen-states.)
The above eigen-states in the 1st Landau gauge are not normalizable states, since the
corresponding wave functions in the $x$-direction are non-normalizable plane-wave
states. However, if~we replace these plane-wave states by normalizable wave-packet states,
we can convert the 1st Landau gauge eigen-states $\vert \Psi^{(L_1)}_{n, k_x} \rangle$
into normalizable states~\cite{WH2022}. In~the following, the~states $\vert \Psi^{(L_1)}_{n, k_x} \rangle$    
are supposed to represent such normalizable states.
Now, we are prepared to compare the expectation values of the mechanical OAM operator
and those of the pseudo OAM operator in three different gauges. It was shown in~\cite{WH2022}
that
\begin{equation}
 \langle \Psi^{(S)}_{n, m} \vert \,\hat{L}^{mech}_z \,\vert \Psi^{(S)}_{n, m} \rangle \ = \ 
 \langle \Psi^{(L_1)}_{n, k_x} \vert \,\hat{L}^{mech}_z \,\vert \Psi^{(L_1)}_{n, k_x} \rangle \ = \  
 \langle \Psi^{(L_2)}_{n, k_y} \vert \,\hat{L}^{mech}_z \,\vert \Psi^{(L_2)}_{n, k_y} \rangle \ = \ 
 2 \,n + 1,
\end{equation}
whereas
\begin{equation}
 \langle \Psi^{(S)}_{n, m} \vert \,\hat{L}^{mech}_z \,\vert \Psi^{(S)}_{n, m} \rangle \ \neq \ 
 \langle \Psi^{(L_1)}_{n, k_x} \vert \,\hat{L}^{mech}_z \,\vert \Psi^{(L_1)}_{n, k_x} \rangle \ \neq \  
 \langle \Psi^{(L_2)}_{n, k_y} \vert \,\hat{L}^{mech}_z \,\vert \Psi^{(L_2)}_{n, k_y} \rangle .
\end{equation}
As anticipated, the~expectation values of the mechanical OAM operator is absolutely 
independent of the gauge choices. In~sharp contrast, the~expectation values of the pseudo 
OAM operators turn out to depend on the choices of gauge. 
An important lesson learned from this analysis is that the gauge-covariant transformation
property of some operator does not necessarily mean the gauge invariance of
the corresponding quantity. The~same can be said for the gauge-covariant
extension of the canonical quark OAM operator in QCD. Despite its gauge-covariance, 
they are not gauge-invariant quantities in a true or traditional sense in perfect
accordance with the insight shown by Ji, Xu, and~Zhao~\cite{JXZ2012}.

\vspace{1mm}
Exactly the same can be said for the gluon spin operator. It is long known that
there is no gauge-invariant local expression of the gluon spin operator.
To resolve the inconsistency between the gauge-invariance issue and its
observability, Ji reopened his longstanding claim that the gluon spin $\Delta G$ has
a meaning only in the light-cone gauge, and~it is not a gauge-invariant quantity 
in a true or traditional sense, although~it is measurable in 
deep-inelastic-scattering processes. One might feel a small self-contradiction in this statement.
This is because we know that the famous gauge principle dictates that gauge non-invariant
quantities do not correspond to observables. Although~most experts tend to avoid 
touching upon such an academic issue, we nevertheless think it is unavoidable to form 
a common recognition, because~it concerns our final shared consensus on the gauge-invariant
nucleon spin decomposition problem. A~likely answer implicitly accepted by several
experts of perturbative QCD would be the following.
That is, the~gluon spin $\Delta G$ is not a genuine observable, but~it is a quasi-
observable, or~to put it more clearly, a~theoretical-scheme-dependent observable. 
This is not so unconventional statement if one remembers the fact that
the genuine observables in the DIS processes are the structure functions not the parton
(quark and gluon) distribution functions. The~latter are recognized as the quantities which
depend on the choice of the regularization or factorization scheme within the
framework of perturbative QCD. To~avoid misunderstanding, we emphasize that, 
different from the gluon spin $\Delta G$, the~quark spin fraction $\Delta \Sigma$ corresponds
to a direct or genuine observable. This is because it can be identified with the flavor--singlet 
axial charge of the nucleon in the gauge-invariant factorization scheme, and~because 
the flavor--singlet axial charge of the nucleon can, in~principle, be observed through the 
neutrino scatterings on the nucleon. It is not the case for the gluon spin, however.
The ultimate reason is that there exist no external electroweak currents, which directly couple 
to the flavor--blind gluon~fields.

\section{Nucleon Generalized Form Factors and Ji's Angular Momentum Sum~Rule}
\label{sec:GPD}

As is well-known, the~electromagnetic form factors of the nucleon are defined
as a non-forward nucleon matrix element of the electromagnetic current $J^\mu$ as
\begin{equation}
 \langle N(P^\prime) |\, J^{\mu} \,| N(P) \rangle
 \ = \ \bar{U} (P^{\prime}) 
 \left[ A_{10} (t) \,\gamma^\mu + 
 B_{10} (t) \, 
 \frac{i \sigma^{\mu \nu} \Delta_{\nu}}{2 M} 
 \right] U (P),
\end{equation}
where $t = (P^\prime - P)^2$, and~$A_{10} (t)$ and $B_{10} (t)$, usually denoted as 
$F_1 (t)$ and $F_2 (t)$, correspond to the familiar Dirac and Pauli form factors
of the nucleon. 
These electromagnetic form factors can be extracted through 
elastic scatterings of the electron from the target nucleon.
On the other hand, the~so-called generalized form factors of the nucleon
are defined as a non-forward nucleon matrix element of the 2nd rank
energy--momentum tensor $T^{\mu \nu}$ as
\begin{equation}
 \langle N(P') |\, T^{\mu \nu}_{q, G} \,| N(P) 
 \rangle \ = \ \bar{U} (P') \left[ \, 
 A_{20}^{q, G} (t) \,
 \gamma^{( \mu} P^{\nu )} \ + \ 
 B_{20}^{q, G} (t) \,
 \frac{P^{( \mu} i \sigma^{\nu ) \alpha} 
 \Delta_\alpha}{2 M} \right] \,U(P).
\end{equation}
(Note that the energy--momentum tensor couples to both of quarks and gluons, while
the electromagnetic current couples only to quarks since gluons are electrically neutral.)
Here, $A_{20} (t)$ and $B_{20} (t)$ are called the gravitational form factors of
the nucleon. As~a matter of course, it is impractical to get information about these
form factors directly through the graviton--nucleon scattering process,
because the gravitational interaction is far weaker than the electroweak
interactions. 
Fortunately, these generalized form factors of nucleons are known to be related to
the quantities called the generalized parton distribution functions, which can 
in principle be extracted through the high-energy scattering processes called the
deeply virtual Compton~scatterings.

\vspace{1mm}
The deeply virtual Compton scattering (DVCS) is the high-energy scattering process 
in which the initial photon is not a real photon but a virtual photon exchanged 
between the projectile lepton and the target nucleon. 
In the Bjorken limit, the~DVCS scattering amplitudes are known to depend on 
four generalized parton distributions (GPDs), $H(x, \xi, t)$, $E(x, \xi, t)$,
and $\tilde{H} (x, \xi,t)$, and~$\tilde{E} (x, \xi, \,t)$, which contain the information
about the nonperturbative quark--gluon structure of the nucleon.
Here, the~first two are called the unpolarized GPDs, while the last two are called
the longitudinally polarized GPDs.
These GPDs all depend on three kinematic variables, $x$, $\xi$, and~$t$. 
The meaning of these kinematic variables is as follows. First, $t$ is 4-momentum
transfer square of the nucleon. Second, $x$ is the average longitudinal momentum 
fraction of the struck quark in the initial and final states, which is sometimes called
the generalized Bjorken variable. Finally, $\xi$ is the difference of the 
longitudinal momentum fractions of the initial and final partons. It is usually called
the skewness~parameter. 

\vspace{1mm}
Of our main interest here is the unpolarized GPDs $H (x,\xi, t)$ and $E (x, \xi, t)$, 
defined as
\begin{eqnarray}
 \int \hspace{-1mm}&\,&\hspace{-4mm} \frac{d \lambda}{2 \,\pi} \, \langle p^\prime, 
 s^\prime \,\vert \, \bar{\psi} \left(- \,\frac{\lambda \,n}{2} \right) \,\slashed{n} \,
 \psi \left(\frac{\lambda \,n}{2} \right) \,\vert p, s \rangle \nonumber \\
 &=& \ \bar{U} (p^\prime, s^\prime) \,\left[ H (x,\xi, t) \,\slashed{n} \ + \ E (x, \xi, t) \,
 \frac{\ i \,\sigma^{\mu \nu} \,n_\nu \,\Delta _\nu}{2 \,M} \,\right] \,
 U (p, s^\prime), \hspace{8mm}
\end{eqnarray}
where $n^\mu$ stands for the familiar light-like 4-vector.
Here, the~so-called gauge link, which is necessary for the above expression to be
gauge-invariant, is omitted for simplicity. It is known that the decomposition
of the nucleon spin is most conveniently made in the Breit frame. In~this reference frame,
the following combination of GPDs $H$ and $E$ naturally appear in the cross-section
formulas, which we denote as $H_E$ and $E_M$: 
\begin{eqnarray}
 H_E (x, \xi, t) \ &\equiv& \ H (x, \xi, t) \ + \ 
 \frac{t}{4 \,M^2_N} \,E (x, \xi, t), \\
 E_M (x, \xi, t) \ &\equiv& \ H (x, \xi, t) \ + \ E (x,\xi, t). 
\end{eqnarray}
This decomposition precisely corresponds to the standard Sachs decomposition of
the nucleon electromagnetic form factors given as
\begin{eqnarray}
 G_E (t) \ &\equiv& \ F_1 (t) \ + \ \frac{t}{4 \,M^2_N} \,F_2 (t), \\
 G_M (t) \ &\equiv& \ F_1 (t) \ + \ F_2 (t).
\end{eqnarray}

We first recall that the sum of $H$ and $E$ satisfy the 1st moment sum rule
given as
\begin{eqnarray}
 \int_{- \,1}^1 \,\left[ H^q (x, 0, 0) \ + \ E^q (x, 0, 0) \right] \,d x &=&
 A^q_{10} (0) \ + \ B^q_{10} (0).
\end{eqnarray}
Here, $A^q_{10} (t)$ and $B^q_{10} (t)$ represent the contributions of a quark with 
flavor $q$ to the familiar Dirac and Pauli form factors of the nucleon.
In the forward limit $t \rightarrow 0$, their sum just gives the contribution of a quark 
with flavor $q$ to the total nucleon magnetic moment (it is the sum of canonical part
and the anomalous magnetic part).

\vspace{1mm}
More interesting to us is the 2nd moment sum rules of $H$ and $E$.
They are given as
\begin{eqnarray}
 \int_{- \,1}^1 \,x \,[\, H^q (x, \xi, t)  \ + \ E^q (x, \xi, t) ] \,d x 
 \ &=& \ A^q_{20} (t) \ + \ B^q_{20} (t), \\
 \int_0^1 \,x \, [\, H^G (x, \xi, t) \ + \ E^G (x,\xi, t) ] \, d x 
 \ &=& \ A^G_{20} (t) \ + \ B^G_{20} (t).
\end{eqnarray}
Here, $A^q (0)$ and $A^G (0)$ respectively correspond to the momentum fraction 
carried by quarks with flavor $q$ and gluons inside the nucleon as
\begin{eqnarray}
 A^q_{20} (0) \ &=& \ \ \int_0^1 \,x \,\left[ q (x) \ + \ \bar{q} (x) \right] \, d x 
 \ \ \equiv \ \ \langle x \rangle^q, \\
 A^G_{20} (0) \ &=& \ \int_0^1 \,x \,g (x) \,d x \ \ \equiv \ \ \langle x \rangle^G. 
\end{eqnarray}
On the other hand, $B^q_{20} (0)$ and $B^G_{20} (0)$ are interpreted as quark 
and gluon contributions to the nucleon anomalous gravito-magnetic moment (AGM).
Different from $A^q_{10} (0)$ and $A^G_{10} (0)$, there is no experimental information 
available for $B^q_{20} (0)$ and $B^G_{20} (0)$ at present. 
Nonetheless, any information for them is strongly desired.
The reason is that they are the quantities which appear in the famous nucleon spin 
sum rule proposed by Ji~\cite{Ji1997, Ji1998}, which is represented as
\begin{equation}
 \frac{1}{2} \ = \ J^Q \ + \ J^G, 
\end{equation}
where
\begin{equation}
 J^Q \ = \ \frac{1}{2} \,\left[\, \langle x \rangle^Q \ + \ B^Q_{20} (0) \,\right], \ \ \ 
 J^G \ = \ \frac{1}{2} \,\left[\, \langle x \rangle^G \ + \ B^G_{20} (0) \,\right] ,
\end{equation}
with the constraint
\begin{equation}
 \langle x \rangle^Q \ + \ \langle x \rangle^G \ = \ 1, \ \ \ 
 B^Q_{20} (0) \ + \ B^G_{20} (0) \ = \ 0 . \label{Eq:constraint}
\end{equation}
Here, $Q$ denotes the sum of all active quark flavors, i.e.,
$Q = u + d + \cdots$. ($Q = u + d$ for the two flavor case, and~$Q = u + d + s$
for the three flavor case).
Then, $\langle x \rangle^Q$ represents the net momentum
fraction carried by all the quarks in the nucleon. On~the other hand, $\langle x \rangle^G$
represents the momentum fraction carried by the gluon in the nucleon.
The equation $\langle x \rangle^Q \ + \ \langle x \rangle^G \ = \ 1$ is nothing but
the familiar longitudinal momentum sum rule of the nucleon, while the relation
$B^Q_{20} (0) + B^G_{20} (0) = 0$ gives the consistency condition for the Ji's sum rule
to hold. Since the momentum fractions $\langle x \rangle^Q$ and $\langle x \rangle^G$
are empirically well-determined by now, we realized that $B^Q_{20} (0) = - \,B^G_{20} (0)$
is only one unknown parameter in Ji's nucleon spin sum~rule.

\vspace{1mm}
Because of the hardness of the experimental GPD analyses, we do not have any reliable 
empirical information about $B^Q_{20} (0)$ or $B^G_{20} (0)$ yet. However,
there are some theoretical challenges in estimating their magnitudes, based on the lattice
QCD simulations and also based on the CQSM. It is very interesting to compare 
the predictions of these two theoretical analyses. Unfortunately, the~estimates within the 
Lattice QCD are given only for fairly large (fictitious) pion mass and the chiral extrapolation 
to the physical pion mass seems to have fairly large uncertainties. 
In view of these circumstances, we tried to modify the effective Lagrangian
of the CQSM to include the arbitrary pion mass parameter in the following 
manner~\cite{WN2006, WN2008}:
\begin{equation}
 {\cal L}_{CQSM} \ = \ {\cal L}_0 \ + \ {\cal L}^\prime,
\end{equation}
where
\begin{eqnarray}
 {\cal L}_0 \ &=& \ \bar{\psi} (x) \,\left(i \,\partial \ - \ M \,U^{\gamma_5} (x) \right) \,\psi (x), \\
 {\cal L}^\prime \ &=& \ \frac{1}{4} \,f^2_{\pi} \,m^2_\pi \,\mbox{tr}_f \,
 \left[ U (x) \ + \ U^\dagger (x) \ - \ 2 \,\right].
\end{eqnarray}
The strategy is that, after~obtaining self-consistent soliton solutions with several values of 
$m_\pi$, we subsequently evaluate desired nucleon observables with the use of these~solutions.

\vspace{-3pt}
\begin{figure}[htpb]
\includegraphics[width=14.5cm]{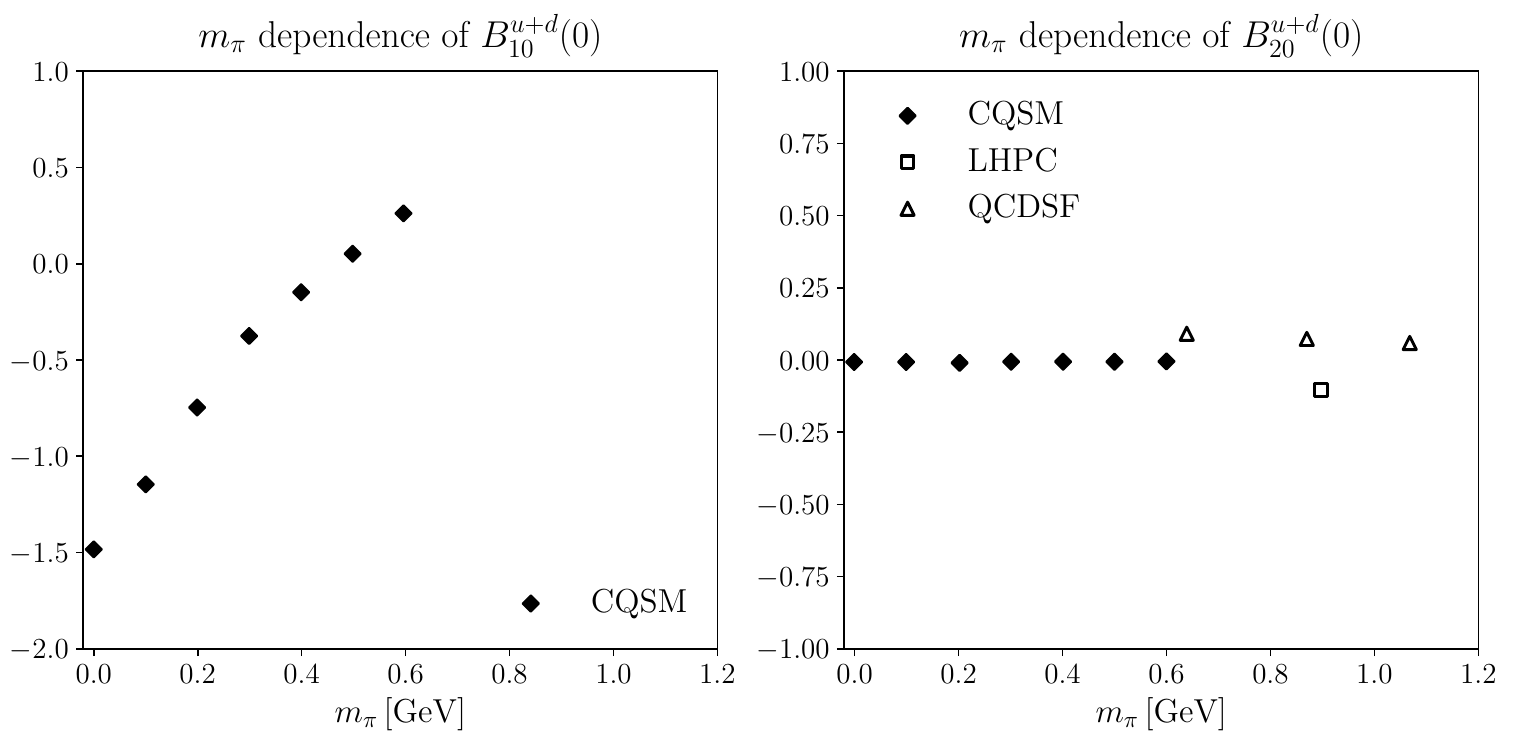}
\caption{The left panel shows the pion mass dependence of  $B^{u+d}_{10} (0)$ 
predicted by the CQSM~\cite{WN2006, WN2008}.
On the other hand, the~right panel shows the pion mass dependence of $B^{u+d}_{20} (0)$ 
predicted by the LHPC and QCDSF lattice simulations carried out with virtually large 
pion masses~\cite{HNRSLS2005, GHHPRSSS2004}. Also shown there is the
prediction of the CQSM for the same quantity.}
\label{fig16}
\end{figure}   

The left panel of Figure~\ref{fig16} shows the pion mass dependence of $B^{u+d}_{10} (0)$
predicted by the SU(2) CQSM~\cite{WN2006, WN2008}. 
Remember that the value of $B^{u+d}_{10} (0)$ is related to the
anomalous magnetic moments of the nucleon as 
$B^{u+d}_{10} (0) = \kappa^u + \kappa^d = 3 \,(\kappa^p + \kappa^n)$,
where $\kappa^u$ and $\kappa^d$ stand for the anomalous magnetic moments of
the $u$-quark and the $d$-quark, while 
$\kappa^p$ and $\kappa^n$ represent the anomalous magnetic moment
of the proton and the neutron, respectively. It is known that this isoscalar combination 
of the nucleon anomalous magnetic moment takes a small negative value
as $\kappa^p + \kappa^n \simeq - \,0.12$. The~predictions of the CQSM seem qualitatively
consistent with this observation.
On the other hand, the~right panel of the same figure shows the pion mass dependence
of $B^{u+d}_{20} (0)$ predicted by the LHPC and QCDSF lattice QCD groups~\cite{HNRSLS2005, GHHPRSSS2004} in comparison with the predictions of
the CQSM. The~CQSM predicts that $B^{u+d}_{20} (0)$
is identically zero independently of the pion mass. It is only natural, since there is a
general constraint such that $B^{u+d}_{20} (0) + B^G_{20} (0) = 0$ and the CQSM
contains no explicit gluon degrees of freedom, thereby indicating that $B^G_{20} (0)$
is identically zero in this effective quark model. Note, however, that the lattice QCD
simulations carried out at larger pion masses also predict a fairly small value of
$B^{u+d}_{20} (0)$, although~it is uncertain at the present stage whether this tendency 
persists even down to the physical pion mass around $138 \,\mbox{\rm MeV}$.
In any case, it should be kept in mind that, once the value of $B^Q_{20} (0)$ or $B^G_{20} (0)$ is 
known, one can decompose the net nucleon spin of one-half into the sum of the total 
quark angular momentum $J^Q$ and the total gluon angular momentum $J^G$.
Besides, one important piece of information obtained from the analysis in the present section is 
that the magnitude of $B^Q_{20} (0)$ or $B^G_{20} (0)$ is likely to be fairly small. 
In the following section, further decomposition of the net nucleon spin will be tried
by making use of this observation, together with some other empirical~information.

\section{Semi-Empirical Analysis of Nucleon Spin Contents and Their Scale~Dependencies}
\label{sec:Nspin}

We again start with the nucleon spin sum rule of Ji given as~\cite{Ji1997, Ji1998}
\begin{equation}
 \frac{1}{2} \ = \ J^Q \ + \ J^G, 
\end{equation}
where
\begin{equation}
 J^Q \ = \ \frac{1}{2} \,\left[\, \langle x \rangle^Q \ + \ B^Q_{20} (0) \,\right], \ \ \ 
 J^G \ = \ \frac{1}{2} \,\left[\, \langle x \rangle^G \ + \ B^G_{20} (0) \,\right] ,
\end{equation}
with the constraint $\langle x \rangle^Q \ + \ \langle x \rangle^G \ = \ 1$ and 
$B^Q_{20} (0) \ + \ B^G_{20} (0) \ = \ 0$.
%
%
As is well-known, momentum fractions $\langle x \rangle^Q$ and 
$\langle x \rangle^G$ are both scale-dependent quantities. 
An interesting observation made by Ji is that $J^Q$ and $J^G$ obey exactly the
same evolution equations as $\langle x \rangle^Q$ and $\langle x \rangle^G$ do.
According to him, this follows from the fact that the~operation which makes the 
angular momentum operator different 
 from the energy momentum operator 
does not change the short distance singularity of the operator.
(See~\cite{Ji1997} for more detail.)
The solution of this (coupled) evolution equation for $(\langle x \rangle^Q, \langle x \rangle^G)$
or $(J^Q, J^G)$ is extremely simple at the leading order (LO) of the perturbative renormalization 
group scheme. It is given as
\begin{eqnarray}
 J^Q (Q^2) \ &=& \ \frac{3 \,n_f}{16 + 3 \,n_f} \, + \, 
 \left(\frac{\ln Q^2_0 / \Lambda^2}{\ln Q^2 / \Lambda} 
 \right)^{2 \,(16 + 3 \,n_f) / (33 - 2 \,n_f)} \,
 \left[ \, J^Q (Q^2_0) \, - \, \frac{16}{16 + 3 \,n_f} \right], \ \ \ \ \ \\
 J^G (Q^2) \ &=& \ \frac{3 \,n_f}{16 + 3 \,n_f} \ + \ 
 \left(\frac{\ln Q^2_0 / \Lambda^2}{\ln Q^2 / \Lambda} 
 \right)^{2 \,(16 + 3 \,n_f) / (33 - 2 \,n_f)} \,
 \left[ \, J^G (Q^2_0) \, - \, \frac{16}{16 + 3 \,n_f} \right], \ \ \ \ \ 
\end{eqnarray}
with $n_f$ being the number of active quark flavor.
In our actual analysis below, we take account of the scale dependencies of the relevant 
quantities by using more involved evolution equations for the momentum fractions at the 
next-to-leading order (NLO) by making use of the previously mentioned  fact
that $(J^Q, J^G)$ (and also $(\langle x \rangle^Q, \langle x \rangle^G)$
obey the same evolution equations. 
As initial data of evolution, we use the MRST fit for the quark and gluon momentum 
fractions~\cite{MRST2004} given
at $Q^2 = 4 \,\mbox{\rm GeV}^2$, 
\begin{equation}
 \langle x \rangle^Q \ = \ 0.579, \ \ \ \langle x \rangle^G \ = \ 0.421.
\end{equation}
Since the lattice QCD indicated that $B^Q_{20} (0) = - \,B^G_{20} (0)$ is likely to be
small, let us simply assume that
\begin{eqnarray}
 J^Q \ &=& \ \frac{1}{2} \,\left[ \langle x \rangle^Q \ + \ B^Q_{20} (0) \right] 
 \ \ \simeq \ \ \frac{1}{2} \,\langle x \rangle^Q, \\
 J^G \ &=& \ \frac{1}{2} \,\left[ \langle x \rangle^G \ + \ B^G_{20} (0) \right]
 \ \ \simeq \ \ \frac{1}{2} \,\langle x \rangle^Q .
\end{eqnarray}
This is a drastic postulate, but~we nevertheless think it useful to gain valuable insights
into the scale dependencies of the nucleon spin contents.
In order to further decompose the total angular momentum of quarks and gluon into
the contributions of intrinsic spins and orbital angular momenta, we adopt
the frequently used definitions of the quark and gluon orbital angular momentum (OAM)
given by
\begin{eqnarray}
 L^Q \ &\equiv& \ J^Q \ - \ \frac{1}{2} \,\Delta \Sigma, \\ \label{Eq:L^Q}
 L^G \ &\equiv& \ J^G \ - \ \Delta G. \label{Eq:L^G}
\end{eqnarray}
It is very important to recognize that the net quark OAM $L^Q$ defined as above is the
mechanical OAM not the canonical OAM. 
Also important to recognize is the fact that the gluon OAM $L^G$ defined as above is not 
the canonical gluon OAM appearing in the Jaffe--Manohar decomposition but it is
the sum of the canonical gluon OAM and what we call the 
potential angular momentum \cite{Waka2010, Waka2011}. 
We emphasize that the potential angular momentum is present only for gluons bound 
in the~nucleon.

\vspace{1mm}
Now let us estimate the nucleon spin contents, especially their scale dependencies 
based on the strategy as follows. 
First, we recall that, as~for the quark and gluon momentum fractions, $\langle x \rangle^Q$ 
and $\langle x \rangle^G$, MRST2004 fit~\cite{MRST2004} and CTEQ5 QCD fit~\cite{CTEQ2000}
give almost the same numbers 
in the range between $Q^2 \simeq 4 \,\mbox{\rm GeV}^2$ and $Q^2 \simeq 10 \,\mbox{\rm GeV}^2$. 
For example, their fits give
\begin{equation}
 \langle x \rangle^Q \ \simeq \ 0.578, \ \ \ \langle x \rangle^G \ \simeq \ 0.422
 \ \ \ \mbox{\rm at} \ \ Q^2 \ =\ 4 \,\mbox{\rm GeV}^2.
\end{equation}
To get full decomposition of the nucleon spin, 
we also need the information for the quark and gluon spin fractions, 
$\Delta \Sigma$ and $\Delta G$, at~$Q^2 = 4 \,\mbox{\rm GeV}^2$.
We already know that $\Delta \Sigma$ is almost scale-independent at these energy scales 
and that $\Delta \Sigma$ is around 0.3. The~value of $\Delta G$ around 
$Q^2 \simeq 4 \,\mbox{\rm GeV}^2$ is not still reliably determined.
There are some global fits at $Q^2 = 10 \,\mbox{\rm GeV}^2$.
For example, the~DSSV collaboration gives $\Delta G \simeq 0.361$ \cite{DSSV2009}.
On the other hand, the~recent lattice calculation  at the physical pion 
mass by the CLQCD collaboration predicts
$\Delta G \simeq 0.231$ at $Q^2 \simeq 10 \mbox{\rm GeV}^2$ \cite{CLQCD2025}. 
As a trial choice for the present qualitative
analysis, we choose 

\begin{equation}
 \Delta G \,(Q^2 =4 \,\mbox{\rm GeV}^2) \ = \ 0.25 .
\end{equation}
Now, under~the approximation that $J^Q \simeq \frac{1}{2} \,\langle x \rangle^Q$
and $J^G \simeq \frac{1}{2} \,\langle x \rangle^G$, the~values of $J^Q$ and $J^G$, as well as
$\Delta \Sigma$ and $\Delta G$, are nonetheless 
 prepared at $Q^2 = 4 \,\mbox{\rm GeV}^2$.
We also know the coupled NLO evolution equation for $(J^Q, J^G)$ 
and also for $(\Delta \Sigma, \Delta G)$. After~solving these evolution equations,
we are able to know the values of these four quantities at arbitrary $Q^2$.
This also enables us to get the values of the net quark and gluon OAMs
at any $Q^2$ from the relations
\begin{equation}
 2 \,L^Q (Q^2) \ = \ 2 \,J^Q (Q^2) \ - \ \Delta \Sigma (Q^2), \ \ \ 
 2 \,L^G (Q^2) \ = \ 2 \,J^Q (Q^2) \ - \ 2 \,\Delta G (Q^2). 
\end{equation}
 (One should keep in mind  the caution given below in 
  Equations
~(\ref{Eq:L^Q}) and (\ref{Eq:L^G}) 
as to the physical meaning of these OAMs.)

\begin{figure}[htpb]
\includegraphics[width=15.0cm]{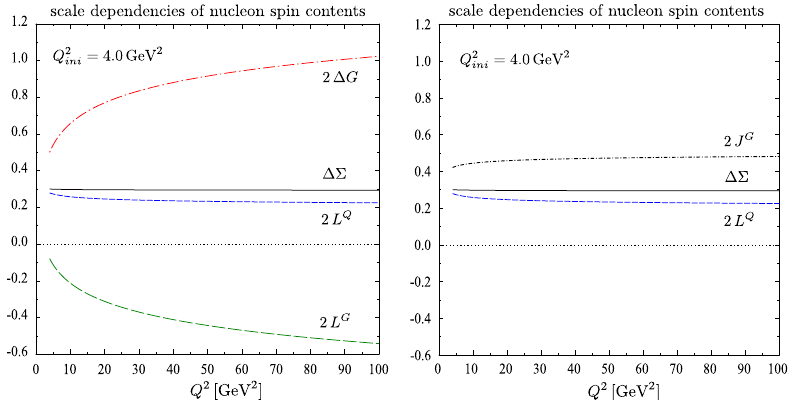}
\caption{The left panel shows the scale dependencies of the four pieces
of the nucleon spin multiplied by two, i.e.,~$\Delta \Sigma$, $2 \,\Delta G$, $2 \,L^Q$, 
and $2 \,L^G$. On~the other hand, on~the right panel, the~net gluon orbital angular 
momentum $2 \,J^Q$ is shown without decomposing it into the spin and OAM~components.}
\label{fig17}
\end{figure}

\vspace{1mm}
We show in Figure~\ref{fig17} the scale dependencies of the nucleon spin contents.
The left panel shows the scale dependencies of the four pieces
of the net nucleon spin multiplied by two, i.e.,~$\Delta \Sigma$, $2 \,\Delta G$, $2 \,L^Q$, 
and $2 \,L^G$. One sees that both the gluon spin and the gluon OAM have
fairly strong scale dependencies even at high energy scales. This means that to~talk about
the decomposition without specifying the energy scale tends to cause confusion.   
On the other hand, on~the right panel of the same figure, the~net gluon angular momentum
$J^G$ is shown without decomposing it to its spin and OAM parts.
In that case, we see that the scale dependence of the three quantities $\Delta \Sigma$, 
$\Delta^Q$, and~$J^Q$ are moderately weak, at least above $Q^2 \simeq 10 \,\mbox{\rm GeV}^2$.   
As pointed out in Section~\ref{sec:GI_decomposition}, in~the original nucleon spin decomposition of Ji, he stated that 
the net gluon angular momentum $J^G$ cannot be gauge-invariantly decomposed into
its intrinsic spin part and the OAM part. Whether this fact has some relation or not with the 
novel observation obtained through the comparison of the left and right panels of Figure~\ref{fig17} 
is a puzzling~question. 

\vspace{-3pt}

\section{Summary and~Outlook}
\label{sec:conclusion}

The CQSM is a unique model of baryons, which has an intimate connection with the 
Skyrme model probably with wider popularity.
Although the former is an effective quark theory while the latter is an effective meson theory,
they share a lot of common features. For~instance, the~classical pion
field configuration of the hedgehog shape in the Skyrme model plays
the role of a mean field for quarks in the CQSM.
The collective quantization of a symmetry-restoring rotational motion of the 
hedgehog object is also a common basic ingredient in both models. 
Despite these strong similarities, a~conspicuous difference between the
two theories was discovered already in the study of familiar low-energy
observables of the nucleon. It is a novel $1 / N_c$ correction, or~more concretely, the~1st-order rotational correction to some observables like the
isovector axial-vector coupling constant $g^{(3)}_A$ of the nucleon, which was found to
exist within the framework of the CQSM but is absolutely missing in the scheme 
of the Skyrme model. Undoubtedly, this $1 / N_c$ correction is an important
key factor that makes the predictions of the CQSM more quantitative than 
those of the Skyrme~model.
 
The superiority or wider applicability of the CQSM over the Skyrme model becomes 
much clearer if the object of study is extended from the low-energy 
observables to the internal partonic structure of the nucleon or any baryons.
Since the parton distribution functions reflect the non-local
light-cone correlation between quarks (and gluons) inside the
nucleon, there is no way to handle them within the framework
of effective meson theories like the Skyrme model.
In contrast, this is just the place where the potential power of the 
CQSM as an effective quark model of baryons manifests itself. 
In fact, we have demonstrated that the CQSM can explain almost all the characteristic 
features of the various types of quark distribution functions of the nucleon.
Among others, worthy of special mention is the fact that the CQSM predictions
for the flavor asymmetry of both the unpolarized and longitudinally polarized 
sea-quark (anti-quark) distributions are remarkably consistent with the empirical 
information obtained from the analyses of the high-energy deep-inelastic~scatterings.

On the other hand, a~weakness of the CQSM is that it does not contain explicit
gluon degrees of freedom, which means that the model is unable to give any meaningful 
predictions for the gluon distribution functions aside from the components incorporated 
through the perturbative QCD evolution.  
To overcome this weakness of the effective quark model, 
the numerical simulations within the lattice QCD have long been hoped for 
as the most promising candidate.
In the past, the~parton distribution functions could not be handled within the framework
of the lattice QCD, since the necessary light-cone quark--quark correlations cannot be
handled within this framework. 
However, a~breakthrough has been made by the advent of the so-called large momentum
effective theory by Ji and collaborators~\cite{Ji2013, JZZ2013}.
In this framework, one first considers the quasi-parton distribution functions, the~evaluation of which requires to handle the space-like quark--quark correlations instead 
of the light-cone correlations. After~that, the~desired genuine parton distribution functions
are constracted from the quasi-parton distributions by means of the sophisticated matching 
procedure via the large momentum nucleon states. 
Based on this strategy, there already exist several challenges to evaluate not only the 
parton distribution functions but also the generalized parton
distribution functions~\cite{Alexandrou2016, Lin2019, Alexandrou2018, Constantinou2021}. 
At the present stage, however, what is mostly calculated are the so-called quasi
parton distribution functions, not the genuine parton distribution functions.
To obtain the genuine parton distribution functions from the quasi-parton distribution functions,
one is required to carry out the sophisticated matching procedure~\cite{Ji2013, JZZ2013}.
It is not necessarily clear whether this matching procedure has already been carried out
at the fully satisfactory level or not.
Besides, we are not completely sure whether the effects of pionic quark--antiquark
excitation modes inside baryons are taken into account at the lattice QCD simulations 
carried out up to now, even though all such effects are in principle believed to be incorporated 
automatically into the simulations within the lattice QCD framework. 
In light of this situation, we believe that the flavor-asymmetries of the sea-quark
distributions will be an important touchstone of whether the theoretical predictions
of the lattice QCD have already reached a fully realistic level or~not.

\vspace{+6pt}
%


\acknowledgments{The 
 author would like to express his deepest gratitude to the members of 
KEK Theory Center, especially to Shunzo Kumano, for~many useful discussions
on the generalized parton distribution functions and many other 
subjects.}

\appendix
\section*{Appendix A. Remarks on the Formally Unified Expression of the Quark and Antiquark~Distributions\label{appA}}
\addcontentsline{toc}{section}{Appendix A}


The familiar quark and antiquark distribution functions are functions of the
Bjorken variable $x$ taking the range $0 < x < 1$. It is sometimes convenient
to formally extend the domain of the variable $x$ into the range $- \,1 < x < 1$.
To explain how it works, we start with the definition of the unpolarized
quark distribution function given as the nucleon matrix element of the
quark bilinear operator with light-cone-separation as follows: 
\begin{eqnarray}
  q (x) \ = \ \int_{- \,\infty}^{\infty} \,\, dz_0 \,\,
  e^{ \,i \,x M_N \,z_0} \,\left. \langle \,N \,\vert \,
  \bar{\psi} (0) \,\gamma^+ \,\psi (z) \,\vert \,
  N \,\rangle \,\right\vert_{\,z^+ = z_\perp = 0} \,,
\end{eqnarray}
with $\gamma^+ = (\gamma^0 + \gamma^3) / \sqrt{2}$.
Here, the~variable $x$ is supposed to take positive values in the
range $0 < x < 1$.
On the other hand, the~unpolarized antiquark distribution is defined as 
\begin{equation}
  \bar{q} (x) \ = \ \int_{- \,\infty}^{\infty} \,\, dz_0 \,\,
  e^{ \,i \,x M_N \,z_0} \,\left. \langle \,N \,\vert \,
  {\bar{\psi}}^c (0) \,\gamma^+ \,{\psi}^c (z) \,
  \vert \,  N \,\rangle \right\vert_{\,z^+ = z_\perp = 0} \, ,
\end{equation}
which is also given in the domain $0 < x <1$. In~the above equation,
$\psi^c$ is the anti-particle field of $\psi$ given as
\begin{equation}
 \psi^c \ = \ C \,\bar{\psi}^T,
\end{equation}
where $C$ is the familiar charge conjugation matrix, while   
$\bar{\psi}^T$ represents the transpose of $\bar{\psi}$.
Using the properties of the charge conjugation matrix, one can
prove the identity
\begin{equation}
 \bar{q} (x) \ = \ - \,q (- \,x) \ \ \ \ \ \mbox{with} \ \ \ \ \ 0 < x < 1.
\end{equation}
This identity dictates that the unpolarized quark distribution with a negative value of 
$x$ is identified with the unpolarized antiquark distribution at a positive value of $x$.
This relation is frequently used, since it helps in evaluating and understanding 
the behavior of the $\bar{d} (x) - \bar{u} (x)$, etc.

\vspace{1mm}
Similarly, the~longitudinally polarized quark and antiquark distributions
are defined~as
\begin{eqnarray}
  q (x) \ &=& \ \int_{- \,\infty}^{\infty} \,\, dz_0 \,\,
  e^{ \,i \,x M_N \,z_0} \, \left. \langle \,N \,\vert \,
  \bar{\psi} (0) \, \gamma^+ \gamma_5 \,\psi (z) \,\vert \,
  N \,\rangle \right\vert_{\,z^+ = z_\perp = 0} \,, \\
  \Delta \bar{q} (x) \ &=& \ \int_{- \,\infty}^{\infty} \,\, dz_0 \,\,
  e^{ \,i \,x M_N \,z_0} \, \left. \langle \,N \,\vert \,
  {\bar{\psi}}^c (0) \,\gamma^+ \gamma_5 \,{\psi}^c (z) \,
  \vert \,  N \,\rangle \right\vert_{\,z^+ = z_\perp = 0} \,, 
\end{eqnarray}
in the range $0 < x <1$. In~this case, on~account of the extra $\gamma_5$ matrix,
we find that
\begin{equation}
 \Delta \bar{q} (x) \ = \ + \, \Delta q (- \,x) \ \ \ \ \ \mbox{with} \ \ \ \ \ 0 < x < 1.
\end{equation}
Although the fact that the quark distributions with negative $x$ is related to
the antiquark distribution with positive $x$ is a property common to
the unpolarized and longitudinally polarized distribution, one must pay
attention to the above sign~difference.






\newpage
\bibliography{TDABbibfile}

\begin{thebibliography}{00}





\bibitem{DPP1988}
Diakonov, D. I.; Petrov, V. Yu.; Pobylitsa, P. V.
A chiral theory of nucleons.
{\em Nucl. Phys. B} {\bf 1988}, {\em 306}, 809.

\bibitem{CBKPWMAG1996}
Christov, C.V.; Blotz, A.; Kim, H.-C.; Pobylitsa, P.; Watabe, T.; Meissner, T.; Ruiz Arriola, E.;  Goeke, K.
Baryons as non-topological chiral solitons.
{\em Prog. Part.  Nucl. Phys.} {\bf 1996}, {\em 37}, 91.


\bibitem{ARW1996}
Alkofer, R.; Reinhardt, H.; Weigel, H.
Baryons as chiral solitons in the Nambu-Jona-Lasinio model.
{\em Phys. Rep.} {\bf 1996}, {\em 265}, 139.

\bibitem{Weigel_Book}
Weigel, H.
\emph{Chiral Soliton Models for Baryons};
{Lecture Notes in Physics}; Springer: Berlin/Heidelberg, Germany, 
 {2008}; Volume {743}.

\bibitem{Witten1983}
Witten, E.
Current algebra, baryons, and quark confinement.
{\em Nucl. Phys. B} {\bf 1983}, {\em 223}, 422; 433




\bibitem{DPPPW1996}
Diakonov, D.; Petrov, V.; Pobylitsa, P.; Polyakov, M.; Weiss, C.
Nucleon parton distributions at low normalization point in the
large $N_c$ limit.
 {\em Nucl. Phys. B} {\bf 1996}, {\em 480}, 341--378.

\bibitem{DPPPW1997}
Diakonov, D.; Petrov, V.; Pobylitsa, P.; Polyakov, M.; Weiss, C.
Unpolarized and polarized quark distributions in the large
$N_c$ limit.
{\em Phys. Rev. D} {\bf 1997}, {\em 56}, 4069--4083.

\bibitem{WGR1997L}
Weigel, H.; Gamberg, L.; Reinhardt, H.
Nucleon structure functions from a chiral soliton.
{\em Phys. Lett. B} {\bf 1997}, {\em 399}, 287--296.


\bibitem{WGR1997}
Weigel, H.; Gamberg, L.; Reinhardt, H.
Polarized nucleon structure functions within a chiral soliton model.
{\em Phys. Rev. D} {\bf 1997}, {\em 55}, 6910--6923.


\bibitem{WK1998}
Wakamatsu, M.; Kubota, T.
Chiral Symmetry and the nucleon structure functions.
{\em Phys. Rev. D} {\bf 1998}, {\em 57}, 5755.

\bibitem{WK1999}
Wakamatsu, M.; Kubota, T.
Chiral Symmetry and the nucleon spin structure functions.
{\em Phys. Rev. D} {\bf 1999}, {\em 60}, 034020.

\bibitem{Waka2003A}
Wakamatsu, M.
Light-flavor sea-quark distributions in the nucleon in the SU(3)
chiral quark soliton model. I. Phenomenological predictions.
{\em Phys. Rev. D} {\bf 2003}, {\em 67}, 034005.


\bibitem{Waka2003B}
Wakamatsu, M.
Light-flavor sea-quark distributions in the nucleon in the SU(3)
chiral quark soliton model. II. Theoretical formalism.
{\em Phys. Rev. D} {\bf 2003}, {\em 67}, 034006.

\bibitem{Ji1997}
Ji, X.
Gauge-invariant decomposition of nucleon spin.
{\em Phys. Rev. Lett.} {\bf 1997}, {\em 78}, 610.




\bibitem{Brodsky2010}
Brodsky, S. J.; Roberts, C. D.; Shrock, R.; Tandy, P. C.
New perspective on the quark condensate.
{\em Phys. Rev. C} {\bf 2010}, {\em 82}, 022201.


\bibitem{KR1984}
Kahana, S.; Ripka, G.
Baryon density of quarks coupled to a chiral field.
{\em Nucl. Phys. A} {\bf 1984}, {\em 429}, 462.

\bibitem{KRS1984}
Kahana, S.; Ripka, G.; Soni, V.
Soliton with valence quarks in the chiral invariant $\sigma$-model.
{\em Nucl. Phys. A} {\bf 1984}, {\em 415}, 351.

\bibitem{WY1991}
Wakamatsu, M.; Yoshiki, H.
A Chiral Quark Model of the Nucleon.
 {\em Nucl. Phys. A} {\bf 1991}, {\em 524}, 561--600.

\bibitem{KWW1999}
Kubota, T.; Wakamatsu, M.; Watabe, T.
Chiral quark soliton model with Pauli-Villars regularization.
{\em Phys. Rev. D} {\bf 1999}, {\em 60}, 014018.

\bibitem{ANW1983}
Adkins, G.S.; Nappi, C.R.; Witten, E.
Static properties of nucleons in the Skyrme model.
{\em Nucl. Phys. B} {\bf 1983}, {\em 228}, 552.

\bibitem{AN1984}
Adkins, G.S.; Nappi, C.R.
Skyrme model with pion mass.
{\em Nucl. Phys. B} {\bf 1984}, {\em 233}, 109.

\bibitem{RS_Book}
Ring, P.; Schuck, P.
\emph{The Nuclear Many-Body Problem};
{Springer: New York, NY, USA; Heidelberg, Germany; Berlin, Germany,} {1980}.

\bibitem{Waka1991}
Wakamatsu, M.
How Dirac-sea quarks affect the neutron charge distribution.
{\em Phys. Lett. B} {\bf 1991}, {\em 269}, 394.



\bibitem{SS2005A}
Sakai, T.; Sugimoto, S.
Low energy hadron physics in holographic QCD.
{\em Prog. Theor. Phys.}  {\bf 2005}, {\em 113}, 843.

\bibitem{SS2005B}
Sakai, T.; Sugimoto, S.
More on a holographic dual of QCD.
{\em Prog. Theor. Phys.}  {\bf 2005}, {\em 114}, 1083.

\bibitem{HSS2008}
Hashimoto, K.; Sakai, T.; Sugimoto, S.
Holographic baryons: Static properties and form factors from gauge/string duality.
{\em Prog. Theor. Phys.} {\bf 2008}, {\em 120}, 1093.

\bibitem{WW1993}
Wakamatsu, M.; Watabe, T.
The $g_A$ problem in hedgehog soliton model revisited.
{\em Phys. Lett. B} {\bf 1993}, {\em 312}, 184.

\bibitem{CBGPPWW1994}
Christov, C.V.; Blotz, A.; Goeke, K.; Pobilitsa, P.; Petrov, V.;
Wakamatsu, M.; Watabe, T.
$1 / N-C$ rotational corrections to $g_A$ and isovector magnetic moment
of the nucleon.
{\em Phys. Lett. B} {\bf 1994}, {\em 325}, 467.

\bibitem{Waka1995}
Wakamatsu, M.
The $g_A$ problem in the Skyrme model and fermion-boson non-correspondence.
{\em Prog. Theor. Phys. Suppl.} {\bf 1995}, {\em 120}, 313.

\bibitem{Waka1996} 
Wakamatsu, M. 
Tracing the origin of the $g_A$ problem in the Skyrme model.
{\em Prog. Theor. Phys.} {\bf 1996}, {\em 95}, 143. 

  
\bibitem{RW2012}
Reinhardt, H.; Weigel, H.
Vacuum nature of the QCD condensates.
{\em Phys. Rev. D} {\bf 2012}, {\em 85}, 074029.

\bibitem{Schweitzer2003}
Schweitzer, P.
The chirally-odd twist-3 distribution function $e (x)$ in the chiral quark-soliton model.
{\em Phys. Rev. D} {\bf 2003}, {\em 67}, 114010.

\bibitem{WO2003}
Wakamatsu, M.; Ohnishi, Y.
Nonperturbative origin of the delta-function singularity in the chirally
odd twist-3 distributio function $e (x)$.
{\em Phys. Rev. D} {\bf 2003}, {\em 67}, 114011.

\bibitem{OW2004}
Ohnishi, Y.; Wakamatsu, M.
$\pi N$ sigma term and chiral-odd twist-3 distribution function $e (x)$
of the nucleon in the chiral quark soliton model.
{\em Phys. Rev. D} {\bf 2004}, {\em 69}, 114002.

\bibitem{Waka2024}
Wakamatsu, M.
Extraordinary nature of the nucleon scalar charge and its densities as
a signal of nontrivial vacuum structure of QCD.
{\em Symmetry} {\bf 2024}, {\em 16}, 1481.

\bibitem{MK1996}
Miyama, M.; Kumano, S.
Numerical solution of $Q^2$ evolution equations in a brute-force method.
{\em Comput. Phys. Commun.} {\bf 1996}, {\em 94},~38.

\bibitem{HKM1998}
Hirai, M.; Kumano, S.; Miyama, M.
Numerical solution of $Q^2$ evolution equations for polarized structure functions.
{\em Comput. Phys. Commun.} {\bf 1998}, {\em 108}, 38.

\bibitem{HERMES1998}
Ackerstaff, K. ~et~al.
Flavor asymmetry of the light quark sea from semi-inclusive 
deep-inelastic scattering.
{\em Phys. Rev. Lett.} {\bf 1998}, {\em 81},~5519.

\bibitem{E866_NuSea1998}
Hawker, E.A.~et~al.
Measurement of the light antiquark flavor asymmetry in the nucleon sea.
{\em Phys. Rev. Lett.} {\bf 1998}, {\em 80}, 3715.


\bibitem{Kumano1998}
Kumano, S.
Flavor asymmetry of antiquark distributions in the nucleon.
{\em Phys. Rep.} {\bf 1998}, {\em 303}, 183.


\bibitem{COMPASS2005}
COMPASS Collaboration; Ageev, E.S.~et~al.
Measurement of the spin structure of the deuteron in the DIS region.
{\em Phys. Lett. B}  {\bf 2005}, {\em 612}, 54.

\bibitem{COMPASS2007}
COMPASS Collaboration; Alexakhin, V.Y.~et~al.
The deuteron spin-dependent structure functoin $g^d_1$ and its first moment.
{\em Phys. Lett. B} {\bf 2007}, {\em 647}, 8--17.

\bibitem{WN2006}
Wakamatsu, M.; Nakakoji, Y.
Generalized form factors, generalized parton distributions, and
the spin contents of the nucleon.
{\em Phys. Rev. D} {\bf 2006}, {\em 74}, 054006.

\bibitem{WN2008}
Wakamatsu, M.; Nakakoji, Y.
Phenomenological analysis of the nucleon spincontents and their scale dependence.
{\em Phys. Rev. D} {\bf 2008}, {\em 77}, 074011.

\bibitem{HERMES2007}
Airapetian, A.~et~al.
Precise determination of the spin structure function $g_1$ of the proton,
deuteron and neutron.
{\em Phys. Rev. D} {\bf 2007}, {\em 75}, 012007.

\bibitem{SMC1998}
Adeva, B.~et~al.
Spin asymmetry $A_1$ and structure functions $g_1$ nof the proton and
the deuteron from polarized high energy muon scattering.
{\em Phys. Rev. D} {\bf 1998}, {\em 58}, 112001.

\bibitem{Waka2009}
Wakamatsu, M.
Transverse momentum distributions of quarks in the nucleon.
{\em Phys. Rev. D} {\bf 2009}, {\em 79}, 094028.


\bibitem{BDGPPP1993}
Blotz, A.; Diakonov, D.; Goeke, K.; Park, N.W.; Petrov, V.; 
Pobylitsa, P.V.
The SU(3) Nambu-Jona-Lasinio soliton in the collective
quantization formulation.
{\em Nucl. Phys. A} {\bf 1993}, {\em 555}, 765.

\bibitem{Sullivan1972}
Sullivan, J.D.
One-pion exchange and deep-inelastic electron-nucleon scattering.
{\em Phys. Rev. D} {\bf 1971}, {\em 5}, 1732.

\bibitem{ST1987}
Signal, A.I.; Thomas, A.W.
Possible strength of the non-perturbative strange sea of the nucleon.
{\em Phys. Lett. B} {\bf 1987}, {\em 191}, 205.

\bibitem{NNPDF2010}
NNPDF Collaboration; Ball, R.D.; Del Debbio, L.; Forte, S.;
Gufanti, A.; Latorre, J. I.; Rojo, J.; Ubiali, M.
A first unbiased global NLO determination of parton distributions and their uncertainties.
{\em Nucl. Phys. B} {\bf 2010}, {\em 832}, 136.

\bibitem{NNPDF2012}
NNPDF Collaboration; Ball, R.D.; Bertone, V.; Cerutti, F.; Del Debbio, L.; Forte, S.;
Gufanti, A.; Latorre, J. I.; Rojo, J.; Ubiali, M.
Unbiased global determination of parton distributions and their uncertainties at
NNLO and at LO.
{\em Nucl. Phys. B} {\bf 2012}, {\em 855}, 153.

\bibitem{BM1996}
Brodsky, S.; Ma, B.-Q.
The quark--antiquark asymmetry of the nucleon sea.
{\em Phys. Lett. B} {\bf 1996}, {\em 381}, 317.

\bibitem{LSS2003}
Leader, E.; Sidorov, A.V.; Stamenov, D.B.
Role of higher twist in polarized deep inelastic scattering.
{\em Phys. Rev. D} {\bf 2003}, {\em 67}, 074017.

\bibitem{LSS2006}
Leader, E.; Sidorov, A.V.; Stamenov, D.B.
Longitudinal polarized parton densities updated.
{\em Phys. Rev. D} {\bf 2006}, {\em 73}, 034023.

\bibitem{DSSV2009}
de Floian, D.; Sassot, R.; Strattmann, M.; Vogelsang, W.
Extraction of spin-dependent parton densities and thir uncertainties.
{\em Phys. Rev. D} {\bf 2009}, {\em 80}, 034030.

\bibitem{SeaQuest2025}
Leung, C.H.~et~al.
inal SeaQuest results on the flavor asymmetry of the proton light-quark
sea with proton-induced Drell-Yan process.
{\em arXiv} {\bf 2025},   arXiv:2512.17564.



\bibitem{JM1990}
Jaffe, R.L.; Manohar, A.
The $g_1$ problem: Deep inelastic electron scattering and
the spin of the nucleon.
{\em Nucl. Phys. B} {\bf 1990}, {\em 337},~509.


\bibitem{Waka2010}
Wakamatsu, M.
Gauge-invariant decomposition of nucleon spin.
{\em Phys. Rev. D} {\bf 2010}, {\em 81}, 114010.


\bibitem{Waka2011}
Wakamatsu, M.
Gauge- and frame-independent decomposition of nucleon spin.
{\em Phys. Rev. D} {\bf 2011}, {\em 83}, 014012.


\bibitem{CLSWG2008}
Chen, X.S.; Lu, X.F.; Sun, W.M.; Wang, F.; Goldman, T.
Spin and orbital angular momentum in gauge theories: Nucleon spin structure 	 
and multipole radiation revisited.
{\em Phys. Rev. Lett.} {\bf 2008}, {\em 100},~232002.

\bibitem{CSLWG2009}
Chen, X.S.; Sun, W.M.; Lu, X.F.; Wang, F.; Goldman,T.
Do gluons carry half of the nucleon momentum?
{\em Phys. Rev. Lett.} {\bf 2009}, {\em 103},~062001.

\bibitem{Hatta2011}
Hatta, Y.
Gluon polarization in the nucleon demystified.
{\em Phys. Rev. D} {\bf 2011}, {\em 84},~041701.

\bibitem{Lorce2013}
Lorce, C.
Gauge-covariant canonical formalism revisited with
application to the proton spin decomposition.
{\em Phys. Rev. D} {\bf 2013}, {\em 88},~044037.

\bibitem{LL2014}
Leader, E.; Lorce, C.
The angular momentum controversy?: What’s it all about 
and does it matter?
{\em Phys. Rep.} {\bf 2014}, {\em 541}, 163.

\bibitem{Waka2014}
Wakamatsu, M.
Is gauge-invariant complete decomposition of the nucleon spin possible?
{\em Int. J. Mod. Phys. A} {\bf 2014}, {\em 29}, 1430012.

\bibitem{JXZ2012}
Ji, X.; Xu, Y.; Zhao, Y.
Gluon spin, canonical momentum, and gauge symmetry.
{\em J. High Energy Phys.} {\bf 2012}, {\em 08}, 082.

\bibitem{BJ1998}
Bashinsky, S.; Jaffe, A.
Quark and gluon orbital angular momentum and spin in hard processes.
{\em Nucl. Phys. B} {\bf 1998}, {\em 536}, 303.

\bibitem{WH2022}
Wakamatsu, M.; Hayashi, A.
Physical symmetries and gauge choices in the Landau problem.
{\em Eur. Phys. J. A} {\bf 2022}, {\em 58}, 121.

\bibitem{WKZ2018}
Wakamatsu, M.; Kitadono, Y.; Zhang, P.-M.
The issue of gauge choice in the Landau problem and the
physics of canonical and mechanical orbital angular momentum.
{\em Ann. Phys.} {\bf 2018}, {\em 392}, 287.

\bibitem{KM2016}
Konstantinou, G.; Moulopoulos, K.
Generators of dynamical symmetries and the correct gauge 
transformation in the Landau level problem: Use of pseudomomentum and 
pseudo-angular momentum.
{\em Eur. J. Phys.} {\bf 2016}, {\em 37}, 065401,

\bibitem{KM2017}
Konstantinou, G.; Moulopoulos, K. 
The “forgotten” pseudomomenta and gauge changes in generalized 
Landau level problems: Spatially nonuniform magnetic and temporally 
varying electric fields.
{\em Int. J. Theor. Phys.} {\bf 2017}, {\em 56}, 1484.



\bibitem{Ji1998}
Ji, X.
Off-forward parton distributions.
{\em J. Phys. G} {\bf 1998}, {\em 24}, 1181.

\bibitem{HNRSLS2005}
H\"agler, P.; Negele, J.W.; Renner, D.B.; Schroers, W.; Lippert, T.; Schilling, K.
Helicity dependent and independent generalized parton distributions of
the nucleon in lattice QCD.
{\em Eur. Phys. J. A} {\bf 2005}, {\em 24}, 29.

\bibitem{GHHPRSSS2004}
G\"ockeler, M.; Hemmert, T.R.; Horsley, R.; Pleiter, D.; Rakow, P.E.L.; Sch\"afer, A.;
Schierholz, G.; Schroers, W.
Nucleon electromagnetic form factors on the lattice and in chiral effective theory.
{\em Nucl. Phys. B Proc. Suppl.} {\bf 2004}, {\em 128}, 203.



\bibitem{MRST2004}
Martin, A.D.; Roberts, R.G.; Stirling, W.J.; Thorne, R.S.
Physical gluons anf high-$E_T$ jets.
{\em Phys. Lett. B} {\bf 2004}, {\em 604}, 61,

\bibitem{CTEQ2000}
Lai, H.L.; Hustopn, J.; Kuhlmann, J.; Olness, F.; Pumplin, J.; Tung, W.K.
Global QCD analysis of parton structure of the nucleon:  CTEQ5 parton distributions.
{\em Eur. Phys. J. C} {\bf 2000}, {\em 12}, 375.

\bibitem{CLQCD2025}
CLQCD Collaboration; Zhao, D.-J.; Chen, L. Dong, H.; Ji, X.; Liu, L.; Pang, Z.;
Sun, P.; Yang, Y.-B.; Zhang, J.-H.;~et~al.
Total gluon helicity contributions to proton spin from lattice QCD.
{\em arXiv} {\bf 2025},  arXiv:2512.243.



\bibitem{Ji2013}
Ji, X.
Parton Physics on a Euclidean Lattice.
{\em Phys. Rev. Lett.} {\bf 2013}, {\em 110}, 262002.

\bibitem{JZZ2013}
Ji, X.; Zhang, J.-H.; Zhao, Y.
Physics of the Gluon-Helicity Contribution to Proton Spin.
{\em Phys. Rev. Lett.} {\bf 2013}, {\em 111}, 112002.

\bibitem{Alexandrou2016}
Alexandrou, C.
Parton distribution functions from lattice QCD.
{\em Few-Body Syst} {\bf 2016}, {\em 56}, 621.

\bibitem{Lin2019}
Lin, H.-W.
Parton distribution functions and lattice QCD.
{\em EPJ Web Conf.} {\bf 2019}, {\em 206}, 01003.

\bibitem{Alexandrou2018}
Alexandrou, C.; Cichy, K.; Constantinou, M.; Jansen, K.; Scapellato, A.;
Steffens, F. 
Light-cone parton distribution functions from lattice QCD.
{\em Phys. Rev. Lett.} {\bf 2018}, {\em 121}, 112001.

\bibitem{Constantinou2021}
Constantinou, M.~et~al.
Parton distributions and lattice-QCD calculations: 
Toward 3D structure.
{\em Prog. Part. Nucl. Phys.} {\bf 2021}, {\em 121}, 103908.


\end{thebibliography}
\bibliographystyle{unsrt}

\end{document}